\documentclass[12pt,a4paper,final]{iopart}

\usepackage{iopams}  
\usepackage{graphicx}
\usepackage[breaklinks=true,colorlinks=true,linkcolor=blue,urlcolor=blue,citecolor=blue]{hyperref}

\usepackage{braket}
\expandafter\let\csname equation*\endcsname\relax
\expandafter\let\csname endequation*\endcsname\relax
\usepackage{amsmath}
\usepackage{float}
\usepackage[toc,page]{appendix}
\usepackage{varwidth}
\allowdisplaybreaks

\begin{document}

\title[Transmission of coherent information  at the onset of interactions]{Transmission of coherent information at the onset of interactions}


\author{Emily Kendall$^{1,2,3}$, Barbara \v{S}oda$^{1,2}$ and Achim Kempf$^{1,2}$} 
\address{$^1$Perimeter Institute for Theoretical Physics, 31 Caroline St N, Waterloo, Ontario, N2L 2Y5, Canada}
\address{$^2$Departments of Applied Mathematics and Physics, and Waterloo Centre for Astrophysics, University of Waterloo, Waterloo, Ontario, N2L 3G1, Canada}
\address{$^3$Department of Physics, The University of Auckland, Private Bag 92019, Auckland, New Zealand}
\ead{emily.kendall@auckland.ac.nz and akempf@uwaterloo.ca}

\begin{abstract}
In this work, we investigate the parameters governing the rate at which a quantum channel arises at the onset of an interaction between two systems, $A$ and $B$. In particular, when system $A$ is pre-entangled with an ancilla, $\tilde{A}$, we quantify the early-time transmission of pre-existing entanglement by calculating the leading order change in coherent information of the complementary channel ($A\rightarrow B'$). 
We show that, when $A$ and $B$ are initially unentangled and $B$ is pure, there is no change in coherent information to first order, while the leading (second) order change is divergent. However, this divergence may be regulated by embedding the conventional notion of coherent information into what we call the family of $n$-coherent informations, defined using $n$-R\'enyi entropies.  
We find that the rate of change of the $n$-coherent information at the onset of the interaction is governed by 
a quantity, which we call the $n$-exposure,  
which captures the extent to which the initial coherent information of $A$ with $\tilde{A}$ is exposed to or `seen by' the interaction Hamiltonian between $A$ and $B$. 
We give examples in qubit systems and in the light-matter interaction.
$$$$

\end{abstract}

%
%
\submitto{\JPA}
%
%
%

\section{Introduction}

Over the past decade, progress in the field of quantum information has seen rapid acceleration, with the physical realisation of a large number of quantum technologies in disciplines such as communication, computing, machine learning, and cryptography. Recently, for example, successful quantum key distribution over distances of thousands of kilometers has been achieved using a ground-to-satellite quantum communication network \cite{Chen2021}. Other recent advances include the development of programmable, scalable photonic quantum chips \cite{Arrazola2021}, and the demonstration of computational performance advantages over classical methods, see, e.g., \cite{King2021}. 

A key challenge in the general field of quantum technologies is to control coherence and entanglement during quantum processes. Depending on the application, different behaviours are desired. Often, it is necessary that entanglement be transferred efficiently from one system to another, such as in quantum communication within or in between quantum processors. Conversely, in other applications it can be important to retain pre-existing entanglement for as long as possible, in particular, to minimize its transfer to the environment \cite{Shor1995, Monz2009}. 

It is therefore important to explore the factors that determine the efficiency with which quantum information is transmitted between two systems as they start interacting. A key tool with which to investigate this is the coherent information \cite{Schumacher1996}. The coherent information serves as an indicator of the degree of preservation of initial entanglement as a quantum system undergoes a quantum channel. 

Moreover, the coherent information is of particular importance in the determination of the capacity of a quantum channel. Specifically, the quantum channel capacity is determined via maximising the coherent information over input states and many parallel uses of the channel \cite{Lloyd1997, Gyongyosi2018, Klesse2007, Cuevas2017}. Via other optimizations, the coherent information also yields the one-way and two-way distillable entanglements. 

In the present work, we analyze the factors that determine the rate at which pre-existing coherent information is transferred at the onset of an interaction. We adopt a perturbative framework, in which we Taylor expand the evolution of the coherent information at early times. Because the coherent information is defined as a difference between von Neumann entropies, we calculate the changes in these entropies perturbatively to yield the overall change in coherent information. 

We find that, while the von Neumann entropies themselves are well-defined for any given channel configuration, the leading order expansion coefficient in their time evolution may diverge, thereby yielding a divergent leading order change in the coherent information. To overcome this problem, we 
extend our analysis by considering a generalisation to a class we refer to as the $n-$coherent information. Whereas the traditional coherent information is defined as a difference of von Neumann entropies, the $n-$coherent information is analogously defined as a difference between the $n^{th}$ R\'enyi entropies. From this perspective, the traditional coherent information may be referred to as the `1-coherent information', since the von Neumann entropy is the limit of the $n^{th}$ R\'enyi entropy as $n\rightarrow 1$.

There are a number of benefits to considering the full family of $n-$R\'enyi entropies rather than only the von Neumann entropy. First, we find that for all integers $n>1$, there are no divergences in the leading order expansion coefficient, and we are therefore able to obtain finite values for the early time behaviour of the corresponding $n-$coherent information. Furthermore, by determining the behavior of the entire family of $n-$R\'enyi entropies we capture a more comprehensive picture of the dynamics. For example, the von Neumann entropy of a density matrix yields little specific information about that density matrix. By contrast, knowledge of the full family of integer R\'enyi entropies for $n>1$ of a density matrix allows one to reconstruct the entire spectrum of the density matrix \cite{Li2008, Flammia:2009axf, PhysRevA.78.032329, 2020PhRvA.102f2413W}. Recall also that the family of R\'enyi entropies with $n>1$ satisfy the additivity and majorisation criteria of suitable information measures \cite{Renyi_og}.

Using the perturbative method, we find that the leading contribution to the time evolution of the $n$-coherent information occurs at second order for all $n$. Moreover, we show that to second order, the evolution of the $n-$coherent information is independent of the free Hamiltonians of either subsystem. This markedly simplifies the computations, and implies that complicated resonance phenomena do not affect the leading order behaviour of the coherent information, since such phenomena necessarily involve the free Hamiltonians in their mathematical descriptions \cite{Hsiang:2019aen, 2022arXiv220108734W}.

Interestingly, we show that the second order change in the $n$-coherent information is not a simple function of the degree of pre-existing entanglement. Instead, it depends upon the extent to which the initial entanglement in the $A\tilde{A}$ system is accessible to the interaction Hamiltonian $H_{int}$ of $A$ with $B$. Specifically, we identify a new quantity, which we term the $n-$exposure, to which the leading order change in the $n-$coherent information is proportional. This is an important point, as it implies that initial preparations of the $\tilde{A}A$ system which possess the same level of entanglement may nevertheless exhibit vastly different behaviours at the onset of interaction with system $B$. Therefore, the notion of $n$-exposure could help, for example, to identify regions in a system's space of states (such as regions in the Bloch sphere in the simple case of a qubit), that are desirable because they have little $n-$exposure and are therefore relatively resistant to transmission of quantum information to the environment. 

To illustrate the utility of the $n-$coherent information and $n-$exposure, we apply our results to simple physical scenarios. We first apply our approach to the quantum Rabi model \cite{rabi1} of the light-matter interaction. In this simple model, we can solve for the time evolution of the qubit exactly. This allows us to verify our perturbative results non-perturbatively. We can then also use the $n-$exposure to identify which regions in the state space have greater propensity to transmit pre-existing entanglement at the onset of interactions. 

We also apply our results to a simple scenario involving a qutrit as the input to a quantum channel in order to explore the utility of the $n-$exposure in larger systems. We explicitly demonstrate regions of state space which are prone to either retaining or transmitting pre-existing entanglement at the onset of the the interaction.   

The structure of this paper is as follows. In Section \ref{sec:prelim}, we prove that free Hamiltonians do not contribute to second order in our calculations and may therefore be neglected in the following sections without loss of generality. We then discuss the notion of coherent information in Section \ref{sec:coherent_information} and describe the quantum channel setup which we will work with throughout the remainder of the paper. In Section \ref{sec:n_1}, we consider the early time evolution of the traditional coherent information, and demonstrate the divergences which can arise in derivatives of the von Neumann entropy. In Section \ref{sec:n_geq_2} we introduce the generalised $n$-coherent information, and demonstrate that, unlike the traditional coherent information, the leading order behaviour is well-defined and finite across state space for $n>1$. In Sections \ref{sec:rabi} and \ref{sec:qutrit} we apply our results to the light-matter interaction and to a qutrit scenario. In Section \ref{sec:conclusion} we discuss the potential significance of our results in the context of quantum communication and computing technologies and address the scope for future work.

\section{Preliminaries}\label{sec:prelim}

In this work, we calculate the transfer of coherent information perturbatively in time, i.e., for the onset of the interaction between two systems. We begin by showing that the transfer of coherent information sets in at second order and that the presence of the free Hamiltonians of the two interacting systems does not contribute to the evolution of the quantities of interest until third order.
This means that, in the remainder of this paper, we will be able to work to leading (i.e., second) order while neglecting the free Hamiltonians. 

To see this, let us start by considering a system $AB$ composed of two subsystems $A$ and $B$, described by density matrix $\rho(t)$, which is separable at $t=0$ such that: $\rho_0 = \rho_A\otimes\rho_B$. We allow the total Hamiltonian  $\hat{H}$ for the time evolution of the combined system $AB$ to be fully general, i.e., of the form: 
\begin{equation}
    \hat{H} := \sum_j \hat{A}_j\otimes\hat{B}_j,
\end{equation}
where the operators $\{\hat{A}_j\}$ act only on subsystem $A$, and the operators $\{\hat{B}_j\}$ act only on subsystem $B$. Note that this general expression for $\hat{H}$  encompasses both interaction terms (with non-trivial action on both subsystems), as well as the free evolution of each subsystem through terms of the form $\hat{A_j}\otimes\mathbb{I}$ (for subsystem $A$) and $\mathbb{I}\otimes\hat{B}_j$ (for subsystem $B$). We now expand the time-evolved state of the total system in the Schr\"odinger picture as: 
\begin{equation}\label{eq:rho_full}
    \rho(t) = e^{it\hat{H}}\rho_o e^{-it\hat{H}}=\rho_0+it\big[\hat{H}, \rho_0\big] + \frac{(it)^2}{2!}\Big[\hat{H}, \big[\hat{H}, \rho_0\big]\Big] + ...
\end{equation}
A key quantity for which we wish to prove that its time evolution is independent of the free Hamiltonians to second perturbative order is the $n$-purity, $\gamma_n$. The $n$-purity of subsystem $B$ is defined as
\begin{equation}
    \gamma_{n,B}(t) := \Tr_B\big[\rho_B(t)^n\big] = \Tr_B\big[\Tr_A[\rho(t)]^n\big], 
\end{equation}
and likewise for subsystem $A$. The first and second time derivatives of the $n$-purity of $B$ are as follows:
\begin{align}
    \dot{\gamma}_{n,B}(t) &= n\Tr_B\Big[\Tr_A[\rho(t)]^{n-1}\Tr_A[\dot{\rho}(t)]\Big],\label{eq:gamma_dot}\\[1em]
    \ddot{\gamma}_{n,B}(t)&=n\Tr_B\Bigg[\Tr_A[\dot{\rho}(t)]\sum_{j=0}^{n-2}\Tr_A[\rho(t)]^j\Tr_A[\dot{\rho}(t)]\Tr_A[\rho(t)]^{n-2-j}\nonumber\\ 
    &\qquad\qquad+\Tr_A[\rho(t)]^{n-1}\Tr_A[\ddot{\rho}(t)]\Bigg].\label{eq:gamma_ddot}
\end{align}
We will first show that, irrespective of the precise form of $\hat{H}$, $\dot{\gamma}_{n,B}(t)$ and $\dot{\gamma}_{n,B}(t)$ vanish at $t=0$. To do this, we use that $\Tr_A[\rho(0)] = \rho_B$, and Equation \ref{eq:rho_full} to find that $\dot{\rho}(0) = i[\hat{H},\rho_0]$. Now, substituting the full expression for $\hat{H}$ we have:
\begin{align}
    \dot{\gamma}_{n,B}(0) &= in\Tr_B\left[\rho_B^{n-1}\Tr_A\Big[\sum_j\hat{A}_j\otimes\hat{B}_j, \rho_A\otimes\rho_B\Big]\right]\nonumber\\[1em]
    &=in\sum_j\Tr_A[\hat{A}_j\rho_A]\Tr_B\big[\rho_B^{n-1}[\hat{B}_j,\rho_B]\big] =0
\end{align}
The expression vanishes because the trace over system $B$ vanishes for every $j$ due to the cyclicity of the trace.

We now consider Equation \ref{eq:gamma_ddot} at $t=0$. Again we note that $\Tr_A[\rho(0)] = \rho_B$ and from Equation \ref{eq:rho_full} we have $\ddot{\rho}(0) = -\big[\hat{H},[\hat{H},\rho_0]\big]$. Hence, \begin{equation}
    \ddot{\gamma}_{n,B}(0) = -n\Tr_B\Bigg[\Tr_A\big[[\hat{H},\rho_0]\big]\sum_{j=0}^{n-2}\rho_B^j\Tr_A\big[[\hat{H},\rho_0]\big]\rho_B^{n-2-j} + \rho_B^{n-1}\Tr_A\big[\hat{H},[\hat{H},\rho_0]\big]\Bigg].
\end{equation}
Substituting in the general expression for $\hat{H}$ and simplifying we find:
\begin{align}\label{eq:full_2nd}
    \ddot{\gamma}_{n,B}(0)=-2n\sum_{jk}\Bigg(&\Tr_A\big[\hat{A}_j\rho_A\big]\Tr_A\big[\hat{A}_k\rho_A\big]\Tr_B\big[\rho_B^{n-1}[\hat{B}_j,\rho_B]\hat{B}_k\big]\nonumber\\
    &+\Tr_A\big[\hat{A}_j\hat{A}_k\rho_A\big]\Tr_B\big[\rho_B^{n-1}[\rho_B\hat{B}_j,\hat{B}_k]\big]\Bigg).
\end{align}
To prove that the free Hamiltonians do not contribute here, we must show that any terms of the form $\hat{A}\otimes\mathbb{I}$ or $\mathbb{I}\otimes\hat{B}$ do not contribute to the sum. To this end, let us consider the Hamiltonian $\hat{H} = \sum_j \hat{A}_j\otimes\hat{B}_j$ where the term with index $j=m$ is of the form $\hat{H}_A\otimes\mathbb{I}_B$. We can now divide the sum over $j,k$ in Equation \ref{eq:full_2nd} as follows:
\begin{equation}\label{eq:sum_division}
    \sum_{jk} = \sum_{j,k\neq m} + \sum_{j,(k=m)} + \sum_{k, (j=m)}.
\end{equation}
There is no contribution from $\hat{H_A}\otimes \mathbb{I}_B$ in the first term on the right of the above equation. Hence, we need only consider the two final terms. We have:
\begin{align}
    -2n\sum_{j,(k=m)}\Big(&\Tr_A\big[\hat{A}_j\rho_A\big]\Tr_A\big[\hat{H}_A\rho_A\big]\Tr_B\big[\rho_B^{n-1}[\hat{B}_j,\rho_B]\mathbb{I}_B\big]\nonumber\\
    &+\Tr_A\big[\hat{A}_j\hat{H}_A\rho_A\big]\Tr_B\big[\rho_B^{n-1}[\rho_B\hat{B}_j,\mathbb{I}_B]\big]\Big)\nonumber\\
    -2n\sum_{k,(j=m)}\Big(&\Tr_A\big[\hat{H}_A\rho_A\big]\Tr_A\big[\hat{A}_k\rho_A\big]\Tr_B\big[\rho_B^{n-1}[\mathbb{I}_B,\rho_B]\hat{B}_k\big]\nonumber\\
    &+\Tr_A\big[\hat{H}_A\hat{A}_k\rho_A\big]\Tr_B\big[\rho_B^{n-1}[\rho_B\mathbb{I}_B,\hat{B}_k]\big]\Big).
\end{align}
We see immediately that there is no contribution from these partial sums because each trace over B either includes a commutator with the identity, or is of the form $\Tr_B\big[\rho_B^{n-1}\hat{B}_j\rho_B-\rho_B^n\hat{B}_j\big]$, which is zero by the cyclicity of the trace. Hence, we see that terms of the form $\hat{H}_A\otimes\mathbb{I}_B$ do not contribute to Equation \ref{eq:full_2nd}. Let us now consider terms of the form $\mathbb{I}_A\otimes\hat{H}_B$. Again, we will assume that the term of this form is indexed by m, and split the double sum over $i,j$ as shown in Equation \ref{eq:sum_division}. Again, the first partial sum, which does not include $j=m$ or $k=m$ does not contain any instances of the $\mathbb{I}_A\otimes\hat{H}_B$ term. Looking at the final two partial sums, we have:
\begin{align}
    -2n\sum_{j,(k=m)}\Big(&\Tr_A\big[\hat{A}_j\rho_A\big]\Tr_A\big[\mathbb{I}_A\rho_A\big]\Tr_B\big[\rho_B^{n-1}[\hat{B}_j,\rho_B]\hat{H}_B\big]\nonumber\\
    &+\Tr_A\big[\hat{A}_j\mathbb{I}_A\rho_A\big]\Tr_B\big[\rho_B^{n-1}[\rho_B\hat{B}_j,\hat{H}_B]\big]\Big)\nonumber\\
    -2n\sum_{k,(j=m)}\Big(&\Tr_A\big[\mathbb{I}_A\rho_A\big]\Tr_A\big[\hat{A}_k\rho_A\big]\Tr_B\big[\rho_B^{n-1}[\hat{H}_B,\rho_B]\hat{B}_k\big]\nonumber\\
    &+\Tr_A\big[\mathbb{I}_A\hat{A}_k\rho_A\big]\Tr_B\big[\rho_B^{n-1}[\rho_B\hat{H}_B,\hat{B}_k]\big]\Big).
\end{align}
We now use the fact that $\Tr_A[\mathbb{I}_A\rho_A] = \Tr_A[\rho_A] = 1$, and rename the index $k\rightarrow j$ in the second sum to write 
\begin{align}
    -2n\sum_{j}\Tr_A[\hat{A}_j\rho_A]\Tr_B\Big[\rho_B^{n-1}\Big(&[\hat{B}_j,\rho_B]\hat{H}_B+[\rho_B\hat{B}_j,\hat{H}_B]\nonumber\\
    &+[\hat{H}_B,\rho_B]\hat{B}_j +[\rho_B\hat{H}_B,\hat{B}_j]\Big)\Big]=0,
\end{align}
where again we have made use of the cyclicity of the trace. 

Hence, we have shown that terms of the form $\hat{A}\otimes\mathbb{I}$ and $\mathbb{I}\otimes\hat{B}$ (i.e. free Hamiltonians) do not contribute to either the first or second time derivatives of the $n$-purity, and therefore we can neglect free evolution when working to second perturbative order in the following sections without loss of generality.

\section{Quantum channels and coherent information}\label{sec:coherent_information}

The controlled isolation or transfer of quantum information among quantum systems is of great importance in quantum technologies \cite{Ladd2010, ASPELMEYER2004}. Of particular interest is the degree to which pre-existing quantum correlations with an ancilla system as measured, e.g., by coherent information (or also, e.g., by negativity) are preserved or transmitted under the action of a quantum channel. 

Let us now consider the direct channel described in the introduction, i.e., the channel from the density matrix of system $A$ at the initial time to the density matrix of system $A$ at a later time, namely after $A$ interacted with a system $B$. In this interaction, $A$ may transmit some of its pre-existing quantum correlations with an ancilla, $\tilde{A}$, to system $B$. Therefore, we also consider the complementary channel from the density matrix of system $A$ at the initial time to the density matrix of system $B$ after the onset of the interaction. 

\begin{figure}[!htb]
\centering
\includegraphics[scale = 0.3, trim={0cm 3cm 0cm 1cm}]{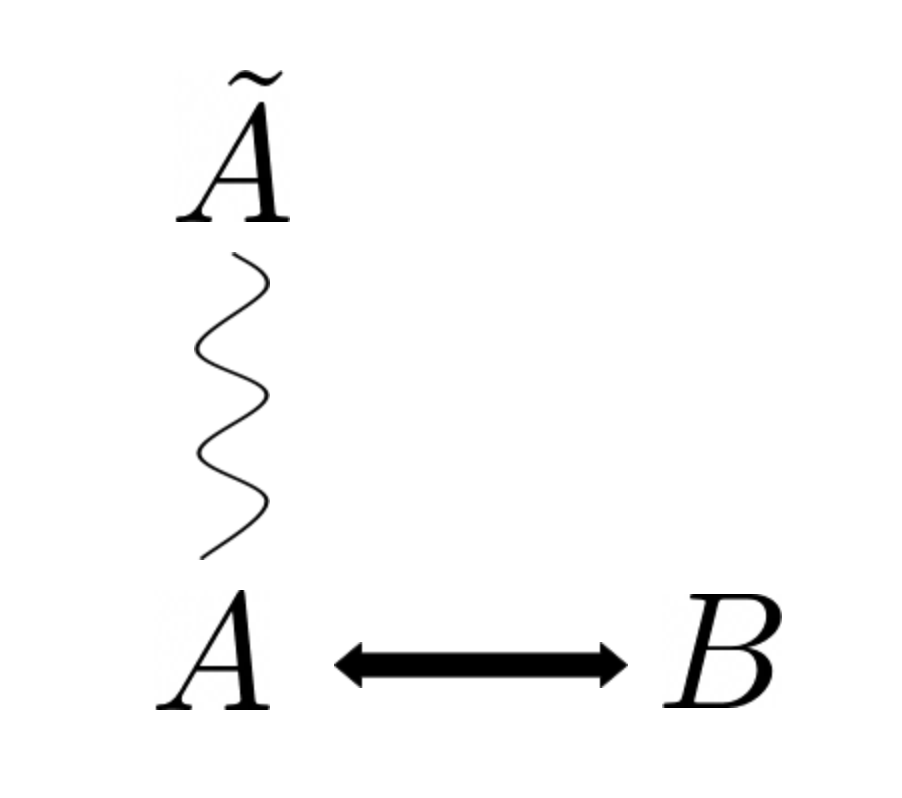}
\caption{A tripartite system in which, initialy, $A$ is purified by $\tilde{A}$, and $B$ is initially pure. An interaction then proceeds between Systems $A$ and $B$ only.}\label{fig:setup}
\end{figure}

We assume that among the three systems, $A$, $\tilde{A}$ and $B$, systems $A$ and $\tilde{A}$ are initially entangled, such that $\tilde{A}$ purifies $A$. $B$ is assumed initially unentangled with both $A$ and $\tilde{A}$, and for now we will assume that system $B$ is initially pure. Therefore, the total tripartite system $A\tilde{A}B$  is also pure. We then consider an interaction which takes place between systems $A$ and $B$ only. This arrangement is illustrated in Figure \ref{fig:setup}.

We consider the direct quantum channel $A\rightarrow A'$ and the complementary channel $A\rightarrow B'$. Here, primes represent the time evolved systems. 
We will study the coherent information of these channels since they are the building blocks for the channel capacity. Moreover, as mentioned before, maximising over all possible input states and parallel executions of the channel gives rise to the overall quantum channel capacity \cite{Lloyd1997, Leditzky2017, Led2018}. 
The coherent information for our two channels is defined as
\begin{align}\label{eq:CI_defn}
    I^d &= S(A') - S((A\tilde{A})'),\\
    I^c &= S(B') - S((B\tilde{A})'),
\end{align}
where the superscripts $d$ and $c$ stand for `direct' and `complementary', respectively, and $S$ represents the von Neumann entropy. 

Though the coherent information is not an entanglement monotone, by definition, the coherent information quantifies the degree to which the entropy of a subsystem exceeds that of its supersystem. Since, classically, this quantity can never be positive, a positive coherent information indicates the presence of quantum correlations between the subsystems \cite{wilde_2017}. 

Since system $B$ starts out pure and system $B\tilde{A}$ starts out mixed, the coherent information, $I^c$, of the complementary channel starts out negative and therefore requires at least a finite amount of time to turn positive. Since $I^c=-I^d$, this means that $I_d$, which starts out positive, needs to decrease. Our aim here is to find out what determines the speed with which the two channel's coherent informations change at the onset of interactions.

\section{Perturbative expansion of the coherent information}\label{sec:n_1}

We now study the evolution of the coherent information perturbatively, calculating the leading order derivatives of the von Neumann entropies in Equation \ref{eq:CI_defn}. We will find that there can be divergences in the leading derivatives, which will motivate the introduction of a generalisation of the coherent information in Section \ref{sec:n_geq_2}.

Let us begin by revisiting the definition of the coherent information, Equation \ref{eq:CI_defn}, which can be re-expressed as follows:
\begin{align}\label{eq:CI_def}
    I^d &= S(A') - S(B'),\\
    I^c &= S(B') - S(A').
\end{align}
Here we have made use of the fact that the entire tripartite system $A\tilde{A}B$ is pure, and therefore any bipartition of the whole must result in a symmetric configuration of entropies. Since the coherent information of the complementary channel is the negative of that of the direct channel, it is sufficient to study the latter.

To compute the leading order change in $I^d$ at the onset of the quantum channel, we note that the von Neumann entropies in Equation \ref{eq:CI_def} can be expressed as the limit of the $n$-R\'enyi entropy, $H_n$, as $n\rightarrow 1$. That is
\begin{equation}
    S(A) = \lim_{n\rightarrow 1} H_{n}(A) = \lim_{n\rightarrow 1} \frac{1}{1-n}\log\big( \gamma_{n,A}\big),
\end{equation}
where $\gamma_{n,A} = \Tr_A[\rho_A^n]$. To obtain the time evolution of the von Neumann entropy, we can therefore compute the evolution of $H_n$ in the limit $n\rightarrow 1$. Thereby we will find that it is not possible to obtain an analytic expression for the leading time derivative of the von Neumann entropy which is valid across the entire state space. 

To see this, let us now consider the case where the interaction Hamiltonian between systems $A$ and $B$ is of the form $H_{int} = \hat{A}\otimes\hat{B}$, generating the unitary time evolution operator $U = \exp\big(it \hat{A}\otimes\hat{B}\big)$. As illustrated in Figure \ref{fig:setup}, systems $A$ and $B$ are assumed initially unentangled with a combined density matrix of the form $\rho_A\otimes\rho_B$. In previous work, see (\cite{KendallKempf}), we already showed that for this situation, the first derivative of the $n$-R\'enyi entropy vanishes for either subsystem, and that the leading (second) order is given by
\begin{equation}\label{eq:deriv_gen}
    \ddot{H}_n(A)|_{t=0} = 
    \frac{\ddot{\gamma}_{n,A}}{(1-n)\gamma_{n,A}}\bigg|_{t=0}=\frac{-2n(\Delta B)^2 \Tr_A\Big[\rho_A^{n-1}\big[\hat{A}, \rho_A\big]\hat{A}\Big]}{(n-1)\Tr_{A}\big[\rho_A^n\big]}, 
\end{equation}
and likewise for system $B$.\footnote{The simplicity of Equation \ref{eq:deriv_gen} highlights the advantage of this perturbative approach. By contrast, in Appendix \ref{app:non-pert} we demonstrate the complexity arising in an exact calculation, from which expressions of this kind cannot be straight-forwardly derived.} In (\cite{KendallKempf}) we assumed the absence of free Hamiltonians of either subsystem. In fact, as we now showed in Section \ref{sec:prelim}, these results hold even with free Hamiltonians as they do not contribute to second order in the evolution of $\gamma$.

In order to obtain the leading derivative of the von Neumann entropy, we must now evaluate Equation \ref{eq:deriv_gen} in the limit $n\rightarrow 1$. We will first attempt to find an analytic expression for the second time derivative as $n\rightarrow 1$. This limit exists, except for states possessing one or more vanishing eigenvalues. To see this, let us first set $n=1+\varepsilon$, and re-express Equation \ref{eq:deriv_gen} accordingly. We have
\begin{equation}\label{eq:epsilon_case}
    \ddot{H}_{1+\varepsilon}(A)|_{t=0} = \frac{-2(1+\varepsilon)(\Delta B)^2 \Tr_A \Big[\rho_A^\varepsilon \big[\hat{A}, \rho_A\big]\hat{A}\Big]}{\varepsilon \Tr_A\big[\rho_A^{1+\varepsilon}\big]},
\end{equation}    
which we may re-express as:
\begin{equation}    
    \ddot{H}_{1+\varepsilon}(A)|_{t=0}=\frac{-2(1+\varepsilon)(\Delta B)^2 \Tr_A \Big[\exp\big(\varepsilon \log \rho_A\big) \big[\hat{A}, \rho_A\big]\hat{A}\Big]}{\varepsilon \Tr_A\big[\rho_A^{1+\varepsilon}\big]}.
\end{equation}
Let us now consider the trace term in the numerator. It can be expanded as a power series in $\varepsilon$:
\begin{equation}
    \Tr_A \Big[\exp\big(\varepsilon \log \rho_A\big) \big[\hat{A}, \rho_A\big]\hat{A}\Big] = \Tr_A \Big[(1 + \varepsilon \log \rho_A + ...) \big[\hat{A}, \rho_A\big]\hat{A}\Big].
\end{equation}
Note that, due to the cyclicity of the trace, the first term in this expansion vanishes since  
\begin{equation}
\Tr_A\Big[\big[\hat{A},\rho_A\big]\hat{A}\Big] = \Tr_A\Big[\hat{A}\rho_A\hat{A}\Big] - \Tr_A\Big[\rho_A\hat{A}\hat{A}\Big] = 0.
\end{equation}
Hence, prior to taking any limits, Equation \ref{eq:epsilon_case} can be written as:
\begin{align}
    \ddot{H}_{1+\varepsilon}(A)|_{t=0} &=\frac{-2(1+\varepsilon)(\Delta B)^2 \Tr_A \Big[\big(\varepsilon \log \rho_A + \mathcal{O}(\varepsilon^2)\big) \big[\hat{A}, \rho_A\big]\hat{A}\Big]}{\varepsilon \Tr_A\big[\rho_A^{1+\varepsilon}\big]},\nonumber\\
    &=\frac{-2(1+\varepsilon)(\Delta B)^2 \Tr_A \Big[\big(\log \rho_A + \mathcal{O}(\varepsilon)\big) \big[\hat{A}, \rho_A\big]\hat{A}\Big]}{\Tr_A\big[\rho_A^{1+\varepsilon}\big]}.
\end{align}
For density matrices $\rho_A$ whose eigenvalues are all nonzero, we obtain an analytic expression for the limit $\varepsilon \rightarrow 0$:
\begin{equation}\label{eq:ddot_H1}
    \ddot{H}_{1}(A)|_{t=0} = -2(\Delta B)^2 \Tr_A \Big[\log \rho_A \big[\hat{A}, \rho_A\big]\hat{A}\Big],
\end{equation}
It can be re-expressed in terms of the eigenvalues ($\lambda_i$) of $\rho_A$: \begin{equation}\label{eq:n_1_case}
    \ddot{H}_{1}(A)|_{t=0} = -2(\Delta B)^2 \sum_{i,j} \log(\lambda_j)(\lambda_i - \lambda_j)\vert a_{ij}\vert^2.
\end{equation}
Taking the limit $n\rightarrow 1$ (or equivalently $\varepsilon \rightarrow 0$) is non-trivial for states $\rho_A$ possessing a vanishing eigenvalue. In fact, if $\rho_A$ possesses a vanishing eigenvalue then we cannot neglect higher order terms in the above expansion, as this would require that the matrix elements of $\varepsilon \log(\rho_A)$ are $ << 1$.
The problem for states with a vanishing eigenvalue, say $\lambda$, arises from the noncommutativity of the two limits $\varepsilon \rightarrow 0$ and $\lambda \rightarrow 0$. To see this, let us consider the trace term in the numerator of Equation \ref{eq:epsilon_case}, expressed in the eigenbasis of $\rho_A$:
\begin{equation}
    \Tr_A \Big[ \rho_A^{\varepsilon} \big[\hat{A}, \rho_A\big]\hat{A}\Big] = \sum_{i,j} \lambda_j^{\varepsilon}(\lambda_i - \lambda_j)\vert a_{ij}\vert^2 = \sum_{i,j} (\lambda_j^{\varepsilon}\lambda_i - \lambda_j^{1+\varepsilon})\vert a_{ij}\vert^2.
    \label{trte}
\end{equation}
The right hand side of the above equation illustrates the issue at hand. Namely, we must consider the term $\lambda^\varepsilon$, as both quantities tend to zero. However, 
\begin{equation}
    \lambda^\varepsilon = \begin{cases}
    0, \text{ for } \lambda = 0, \varepsilon > 0\\
    1, \text{ for } \lambda > 0, \varepsilon = 0
    \end{cases}
\end{equation}
In Appendix \ref{numer}, we numerically illustrate the small $\varepsilon$ and $\lambda$ behavior of the trace term.
The noncommutativity of the two limits $\varepsilon \rightarrow 0$ and $\lambda \rightarrow 0$ is a kind of instability in the sense that dimensions in the Hilbert space which initially have zero probability tend to immediately become populated. An initial divergence does not imply that their corresponding probability increases to a large value, but merely that the rise possesses a large acceleration initially. We discuss this further in Appendix \ref{app:caution}.

To summarize our findings so far: when System $A$ (of Figure \ref{fig:setup}) possesses only finite eigenvalues $\{\lambda_i\}$, we obtain a well defined limit of $\ddot{H}_{1+\varepsilon}(A)\vert_{t=0}$ as $\varepsilon\rightarrow 0$, i.e., for the von Neumann entropy. Conversely, as one or more $\lambda_i\rightarrow 0$, $\ddot{H}_{1+\varepsilon}(A)$ diverges. This includes the case where all eigenvalues but one approach zero, i.e. a pure state. Hence, while the von Neumann entropy itself always approaches a finite limit when one or more eigenvalues tend to zero, its second time derivative does not necessarily do so, see Appendix \ref{app:exact_plots}. 

This fact is important also because, in the setup illustrated in Figure \ref{fig:setup}, system $B$ is initially pure by design, i.e., $\rho_B$ initially possesses some vanishing eigenvalues. Consequently, the leading order change in the coherent information diverges irrespective of the state of System $A$, due to the presence of $\ddot{S}(B)$ in the expression. Indeed, there are a number of further subtleties associated with the perturbative expansion, which we discuss in Appendix \ref{app:caution}. 

In the next section, we therefore introduce a generalisation of the coherent information which is stable as it does not require the taking of the limit $\epsilon\rightarrow 0$.

\section{Generalisation: $n$-coherent information, $n$-durability and $n$-exposure}\label{sec:n_geq_2}

In the previous section we demonstrated that the leading order change in the coherent information suffers a divergence 
(see Figure \ref{fig:setup}), tracing back to the fact that the coherent information is conventionally defined as the difference of von Neumann entropies with the von Neumann entropy being the potentially divergent limit as $n\rightarrow 1$ of the class of $n$-R\'enyi entropies. In this section, we therefore  generalise the notion of coherent information to the entire class of $n$-R\'enyi entropies. In the literature, studies of the wider class of $n$-R\'enyi entropies are greatly increasing in significance due to the usefulness of these entropies in extracting information about the entanglement spectrum of a quantum system,  providing richer information than the von Neumann entropy alone \cite{Islam2015, Li2008}. We note also that work is ongoing in developing novel entanglement measures from the R\'enyi entropies
\cite{Song2016, San2010, Wang2016}.

\subsection{The $n$-coherent information}
We now define the class of $n$-coherent informations, $I_n$, based on the $n$-R\'enyi entropy, through
\begin{align}\label{eq:n_Ic}
    I_{n}^d &= H_n(A') - H_n((A\tilde{A})') = H_n(A') - H_n(B') ,\\
    I_{n}^c &= H_n(B') - H_n((B\tilde{A})') = H_n(B') - H_n(A'),
\end{align}
where the superscripts $d$ and $c$ refer to the direct and complementary channels, respectively. In particular, we have: 
\begin{equation}
    I^d_n=-I^c_n \label{sinv}
\end{equation}
To see this, recall that the two reduced density matrices of two subsystems making up a pure system possess the same nonzero eigenvalues including their multiplicities. 
The $1$-coherent information is the traditional coherent information based on the von Neumann entropy. 

As we will now show, for integer $n>1$ the leading order change of the $n$-coherent information remains well-defined across the whole state space of System $A$, in contrast to the $1$-coherent information, thereby serving as a regulator. Furthermore, as we will discuss, the $n$-R\'enyi entropies may themselves serve as measures of quantum correlations. 

Let us recall here that the $n$-R\'enyi entropy obeys the usual axioms of entropies except for the axiom of subadditivity, except in the cases $n=0$ or $n=1$, see, for example, \cite{RevModPhys.50.221}. This raises the question as to whether the $n$-coherent information for $n\notin\{0,1\}$, besides serving as a regulator for the traditional $1$-coherent information, also serves a direct role in quantum information. Indeed, as was shown in \cite{2002quant.ph..4093V} the $n$-R\'enyi entropies for all $n$ obey an axiom of weak subadditivity: 
\begin{equation}\label{eq:hayden}
    H_n(\rho_A) - H_0(\rho_B) \leq H_n(\rho_{AB}) \leq H_n(\rho_A) + H_0(\rho_B).
\end{equation}
Here, $H_0 = \log(d)$, where $d$ is the Hilbert space dimension of the quantum system, is also called the max-entropy or the Hartley entropy.  
This implies that our notion of $n$-coherent information obeys:
\begin{align}
    H_n(\rho_A) - H_n(\rho_{AB})  \leq H_0(\rho_B)\\
    H_n(\rho_A) - H_n(\rho_{AB}) \geq - H_0(\rho_B)
\end{align}
and therefore:
\begin{equation}
    - H_0(\rho_B) \leq I_n^c \leq  H_0(\rho_B),
\end{equation}
where $I_n^c$ is precisely our notion of the $n-$coherent information. This property of the new notion of $n$-coherent information, in the form of inequality Eq. \ref{eq:hayden}, was shown in \cite{2002quant.ph..4093V} to be crucial for proving a lower bound on quantum communication complexity, namely a lower bound on the number of qubits that two parties need to exchange to be able to turn a specified joint state into another specified joint state. 

By contrast, the traditional ($1$-)coherent information provides a lower bound to the quantum channel capacity. Since the $1$-coherent information is a limiting case of the general $n$-coherent information, this motivates further exploration of the way in which the the quantum channel capacity may be viewed as a limiting case of the above quantum communication complexity. We will revisit this notion in Section \ref{sec:conclusion}.

\subsection{The $n$-durability}

To evaluate the leading order time evolution of the $n$-coherent information, let us recast Equation \ref{eq:deriv_gen} in the form
\begin{equation}\label{eq:definitions}
    \ddot{H}_n(A)|_{t=0} = \frac{2n(\Delta B)^2 D_{n,A}}{(n-1)}, 
\end{equation}
where we defined $D_{n,A}$ as: 
\begin{equation}\label{eq:durability}
    D_{n,A} := -\frac{\Tr_A\Big[\rho_A^{n-1}\big[\hat{A}, \rho_A\big]\hat{A}\Big]}{\gamma_{n,A}}.    
\end{equation}
We will refer to the quantity $D_{n,A}$ as the `$n$-durability', for reasons which will become clear as we progress. 
First, we restrict our attention to integer $n\geq 1$ and we notice that the $n$-durability is a strictly positive quantity. To see this, consider the trace term as expressed in terms of the eigenvalues of the $\rho_A$:
\begin{equation}\label{eq:n_g_1_case}
    \Tr_A \Big[ \rho_A^{n-1} \big[\hat{A}, \rho_A\big]\hat{A}\Big] = \sum_{i,j} \lambda_j^{n-1}(\lambda_i - \lambda_j)\vert a_{ij}\vert^2.
\end{equation}
Noting that $0 \leq \lambda_i\leq 1$, let us consider two eigenvalues, $\lambda_x$ and $\lambda_y$ with $\lambda_x < \lambda_y$. We will therefore have a positive contribution to the sum of the form $\lambda_x^{n-1}(\lambda_y-\lambda_x)\vert a_{xy}\vert^2$. However, we then also have a negative contribution to the sum of the form $-\lambda_y^{n-1}(\lambda_y-\lambda_x)\vert a_{xy}\vert^2$. This negative contribution always outweighs the positive contribution as $\lambda_y^{n-1} > \lambda_x^{n-1}$. Hence, the $n$-durability defined with the negative sign as in Equation \ref{eq:durability} obeys $D_{n,A}\geq 0$.

Not only is the $n$-durability a positive quantity, it also has the useful property that it reduces to the variance in the case of pure states. To see this, we note that if $\rho_A$ is pure then $\rho_A^2 = \rho_A$, and
\begin{align}\label{eq:derive_purity}
    D_{n,A}= -\frac{\Tr_{A}\Big[\rho_A^{n-1}\big[\hat{A},\rho_A\big]\hat{A}\Big]}{\Tr_A[\rho_A^n]} & = -\frac{\Tr_{A}\Big[\rho_A\big[\hat{A},\rho_A\big]\hat{A}\Big]}{\Tr_A[\rho_A]} \nonumber\\
    & = -\Tr_{A}\Big[\rho_A\hat{A}\rho_A\hat{A}\Big] + \Tr_A\Big[\rho_A\hat{A}^2\Big] \nonumber\\
    & = -\sum_i\Big( \bra{i}\ket{\psi}\bra{\psi}\hat{A}\ket{\psi}\bra{\psi}\hat{A}\ket{i} - \bra{i}\ket{\psi}\bra{\psi}\hat{A}^2\ket{i} \Big)\nonumber\\
    & = \bra{\psi}\hat{A}^2\ket{\psi} - \bra{\psi}\hat{A}\ket{\psi}^2\nonumber\\
    & = (\Delta A)^2,
\end{align}
where we defined $\rho_A = \ket{\psi}\bra{\psi}$ and  $\{\ket{i}\}$ is a set of orthonormal basis vectors, one of which can be chosen equal to $\ket{\psi}$. Hence: 
\begin{equation}\label{eq:frag_purity}
    D_{n,A} \xrightarrow{\text{purity}} (\Delta A)^2.
\end{equation}
We note here already that the variance does not constitute a bound on the $n$-durability, as we will explicitly see later.  

Let us now consider the role that the $n$-durability plays in the early time evolution of the $n$-coherent information. The leading order change in the $n$-coherent information of the direct channel ($\delta I_n^d$) is given by
\begin{align}\label{eq:delta_n}
    \delta I_n^d &= \frac{t^2}{2}\left(\ddot{H}_n(A)\big|_{t=0} -\ddot{H}_n(B)\big|_{t=0} \right) \nonumber\\[1em]
    &=\frac{n t^2}{n-1}\Big((\Delta B)^2 D_{n,A}-(\Delta A)^2D_{n,B}\Big)\nonumber\\[1em]
    &=-\frac{n t^2(\Delta B)^2}{n-1}\Big((\Delta A)^2 - D_{n,A}\Big),
\end{align}
Here, we used the fact that System B is pure at $t=0$ such that, through Equation \ref{eq:frag_purity}, $D_{n,B} = (\Delta B)^2$. By the same reasoning, we see that if system $A$ is pure, then $D_{n,A} = (\Delta A)^2$ and $\delta I^d_n =0$, as expected.\footnote{See Appendix \ref{app:caution} for a discussion of the non-trivial limit of Equation \ref{eq:delta_n} as $n\rightarrow 1$.} 

\subsection{The $n$-exposure}
Equation \ref{eq:delta_n} demonstrates that it is not merely the overall amount of quantum correlations that system $A$ initially possesses with the ancilla $\tilde{A}$ which determines the rate at which the $n$-coherent information changes at the onset of interaction. Rather, the $n$-coherent information is sensitive to the amount of system $A$'s entanglement which is `exposed' to the action of interaction Hamiltonian, as determined by the difference between the variance and the $n$-durability. We will therefore refer to this difference as the `$n$-exposure', $E_{n,A}$:
\begin{equation}\label{eq:ee}
    E_{n,A}:=\Big((\Delta A)^2 - D_{n,A}\Big)
\end{equation}

We see from Equation \ref{eq:frag_purity} that in the case that system $A$ is pure, $E_{n,A}$ is zero. This is of course to be expected because if system A does not possess quantum correlations with $\tilde{A}$ to begin with, then no quantum correlations can be transferred during the interaction and the $n$-coherent information should not change. Conversely, if system A possesses quantum correlations with $\tilde{A}$, the exposure is, in general, non-zero. 

However, as we anticipated, the value of the $n$-exposure depends not only on the absolute extent to which System $A$ is entangled or quantum correlated with $\tilde{A}$, but also on the extent to which these quantum correlations are `accessible' or `exposed' to the operator $\hat{A}$ in the interaction Hamiltonian. For illustration, consider the following scenario:

Let system A be comprised of two subsystems, $A_1$ and $A_2$, with a bipartite density matrix of the form $\rho_{A_1}\otimes\rho_{A_2}$. Assume that operator $\hat{A}$ acts only on system $A_1$, such that we may represent the operator on the Hilbert space of $A_1$ and $A_2$ as $\hat{A}_1\otimes\hat{\mathbb{I}}_2$. Let us now evaluate the numerator and denominator of $D_{n,A}$. The numerator is given by:
\begin{align}
     &\Tr_{A_1}\Tr_{A_2}\Big[\rho_{A_1}^{n-1}\otimes\rho_{A_2}^{n-1}\big[\hat{A_1}\otimes\hat{\mathbb{I}}_2, \rho_{A_1}\otimes\rho_{A_2}\big]\hat{A}_1\otimes\hat{\mathbb{I}}_2\Big]\nonumber\\
    &=\Tr_{A_1}\Tr_{A_2}\Big[\rho_{A_1}^{n-1}\big[\hat{A}_1,\rho_{A_1}\big]\hat{A_1}\otimes\rho_{A_2}^n\Big]\nonumber\\
    &=\Tr_{A_1}\Big[\rho_{A_1}^{n-1}\big[\hat{A}_1,\rho_{A_1}\big]\hat{A_1}\Big]\Tr_{A_2}\Big[\rho_{A_2}^n\Big],
\end{align}
while the denominator ($n$-purity) is given by:
\begin{align}
    \gamma_{n,A} &= \Tr_{A_1}\Tr_{A_2}\Big[\rho_{A_1}^n\otimes\rho_{A_2}^n\Big]\nonumber\\
    &=\Tr_{A_1}\Big[\rho_{A_1}^n\Big]\Tr_{A_2}\Big[\rho_{A_2}^n\Big].
\end{align}
Hence, the $\Tr_{A_2}\big[\rho_{A_2}^n\big]$ terms cancel in the numerator and denominator, and we find that 
\begin{equation}
    E_{n,A}=(\Delta A)^2 - D_{n,A_1},
\end{equation}
where, as it should be, the only contribution to the variance comes from system $A_1$. Hence, there even exist extreme cases in which system $A$ overall is highly entangled with system $\tilde{A}$ (from Figure \ref{fig:setup}), but where this entanglement can be entirely accounted for by subsystem $A_2$ while subsystem $A_1$ may not be entangled with $\tilde{A}$ at all. In this case, while system $A$ as a whole is highly entangled with the ancilla $\tilde{A}$, the degree to which that entanglement is `exposed' to the action of the operator $\hat{A}$ is vanishing, since operator $\hat{A}$ has non-trivial action only on subsystem $A_1$. Notice that, due to our results in section \ref{sec:prelim}, remarkably, this conclusion holds true, to second order in time, even in the presence of free Hamiltonians on system $A$, which generically includes an interaction between $A_1$ and $A_2$.

\subsection{Properties of the $n$-exposure}

The $n$-exposure determines the leading order change of the $n$-coherent information of the direct and complementary channels that arise at the onset of the interaction of systems $A$ and $B$. 
Intuitively, the $n$-exposure is the extent to which pre-existing entanglement between $A$ and $\tilde{A}$ is `exposed' to the interaction Hamiltonian of $A$ and $B$. The higher the exposure, the faster the coherent informations of the direct and complementary channels change at the onset of the interaction. 
We found that, in the case of interaction Hamiltonians of the form $\hat{A}\otimes\hat{B}$, the $n$-exposure is the difference between two terms: $E_{n,A}= (\Delta A)^2 -D_{n,A}$, where $(\Delta A)^2$ is the variance in the observable $\hat{A}$ and where  $D_{n,A}$ is the $n$-durability of system $A$. Both are properties of system $A$ only: to calculate the $n$-exposure of system $A$ merely requires knowledge of the operator $\hat{A}$ of the interaction Hamiltonian and the initial reduced density matrix, $\rho_A$. 

For initially pure states of $A$, the $n$-durability equals the variance, i.e., the $n$-exposure vanishes for pure states. 
For initial states of $A$ that are mixed, the $n$-durability can be smaller than the variance. In this case, the $n$-exposure is positive, implying that the $n$-coherent information of the complementary channel rises while, due to Equation \ref{sinv}, that of the direct quantum channel drops at the onset of the interaction. 
Conversely, there also exist initial mixed states of $A$ for which the $n$-durability exceeds the variance. In such cases, the $n$-exposure is negative, and the $n$-coherent information of the complementary channel drops while that of the direct channel rises at the onset of the interaction. 

To further illustrate the significance of negative and positive changes in the $n$-coherent information, let us consider the example of $n=2$. For $n=2$, the $n$-R\'enyi entropy is a simple function of the purity, $\Tr[\rho^2]$,
\begin{equation}
    H_2(A) = - \log(\gamma_{2,A}),
\end{equation}
where $\gamma_2$ is the purity. The purity is useful in that even for mixed bipartite states it is possible to prove that entanglement exists between the subsystems if the purity of one subsystem is lower than the purity of the bipartite system as a whole \cite{Horodecki1996, Islam2015}. In our case, if
\begin{equation}
    \gamma_{2}(A) < \gamma_{2}(\tilde{A}A), 
\end{equation}
or equivalently,
\begin{equation}\label{eq:ineq_2}
    H_{2}(A) > H_{2}(\tilde{A}A), 
\end{equation}
then entanglement must exist between subsystems $A$ and $\tilde{A}$. In fact, the same inequality applies for any $n$-R\'enyi entropy \cite{HORODECKI1996377}. This inequality is known to be a valuable tool to infer the presence of entanglement under non-ideal experimental conditions \cite{Mintert2007}. Furthermore, we note that the $2$-R\'enyi entropy provides a direct lower bound on the von Neumann entropy, while stricter bounds can be obtained using higher order R\'enyi entropies \cite{Daley2012}:
\begin{align}
    H_1(A) &\geq H_2(A),\\
    H_1(A) &\geq 2H_2(A) - H_3(A).
\end{align}
Hence, the $n$-coherent information behaves similarly to the traditional $1$-coherent information in the sense that positivity indicates the presence of quantum correlations. In order, for example, to prevent loss of quantum correlations to an environment system $B$,
it is advisable to avoid input states for which the $n$-coherent information decreases to leading order. This is to prevent the $n$-coherent information from becoming negative, at which point we would no longer fulfill inequality \ref{eq:ineq_2} that guarantees bipartite entanglement between $A$ and $\tilde{A}$. Meanwhile, input states which lead to an increase in the $n$-coherent information are desirable, as they provide a `safe zone' in which the inequality is increasingly fulfilled. 

Therefore, while the $n$-coherent information does not constitute a direct measure of bipartite entanglement, its behavior does provide an indication of which input states of $A$ are more or less vulnerable to losing quantum correlations with $\tilde{A}$ to an environment, $B$. Indeed, in order to obtain the quantum channel capacity from the coherent information, one maximises over all possible input states of the $\tilde{A}A$ system and over multiple parallel uses of the channel. In this optimisation, some states will induce a positive change in the coherent information, while others will lead to a negative change. The preferred input state depends on whether the desired outcome is the transmission or retention of pre-existing quantum correlations between $A$ and $\tilde{A}$.

In the next section, we will apply the notion of $n$-coherent information to the light-matter interaction. We will explore how the leading order behaviour of the $n$-coherent information, as determined by the $n$-exposure, varies with the input state. We will focus on the $n=2$ case for simplicity, but note that a similar analysis can be performed straightforwardly for other $n$.

\section{Application to the light-matter interaction}\label{sec:rabi}

In practice, much of quantum communication relies upon the interaction between light and matter. In this section, we apply our results to a simplified model of the light-matter interaction, namely the quantum Rabi model. At its core, this model describes a two-level quantum system coupled to a single mode of a massless scalar field \cite{rabi1, rabi2, rabi4}. This model is of particular utility as it can be implemented through a variety of experimental techniques such as Josephson junctions \cite{rabi3}, trapped ions \cite{rabi5}, and superconductors \cite{rabi6}.

In the present context, the qubit and the field are chosen initially unentangled and in a product state $\rho_f\otimes\rho_q$. We assume the initial state of the field, $\rho_f$, to be the vacuum state $
    \rho_f =  \ket{0}\bra{0}
$.
We allow the initial state, $\rho_q$, of the qubit to be arbitrary. Expressed in terms of the eigenstates of the $\sigma_z$ Pauli operator it reads
\begin{equation}
    \rho_q = \delta\ket{z^+}\bra{z^+} + \alpha\ket{z^+}\bra{z^-} + \alpha^*\ket{z^-}\bra{z^+} + (1-\delta)\ket{z^-}\bra{z^-},
\end{equation}
where $\sigma_z\ket{z^+} = \ket{z^+}$ and $\sigma_z\ket{z^-} = -\ket{z^-}$. We then consider the interaction of a single field mode with the qubit, governed by an interaction Hamiltonian $H_{\text{int}}$ of the form
\begin{equation}
    H_{\text{int}} = \nu\sigma_z\otimes  (a + a^\dagger)
\end{equation}
where $a$ and $a^\dagger$ are the annihilation and creation operators of the mode considered and where $\nu$ is the effective coupling constant. Since we will neglect the free Hamiltonians, we can set $\nu = 1$ and instead absorb the $\nu$-dependence in the $t$-dependence. We will now solve the problem non-perturbatively in order to then obtain an exact expression for the time evolution of the $n$-R\'enyi entropy, and compare its second time derivative to our Equation \ref{eq:deriv_gen}. Note that while the particularly simple form of the interaction Hamiltonian invoked here allows for a non-perturbative calculation to be performed, this is not always the case. See Appendix \ref{app:non-pert} for details.

We first express the time-evolved system as:
\begin{equation}
    \rho(t) = e^{it\sigma_z\otimes(a+a^\dagger)}(\rho_q\otimes\rho_f) e^{-it\sigma_z\otimes(a+a^\dagger)}.
\end{equation}
For the density matrix of the qubit this gives, see \cite{KendallKempf} for details:
\begin{equation}\label{eq:time_ev_rho}
    \rho_q(t) = \delta\ket{z^+}\bra{z^+} + \alpha e^{-2t^2}\ket{z^+}\bra{z^-} + \alpha^* e^{-2t^2}\ket{z^-}\bra{z^+} + (1-\delta)\ket{z^-}\bra{z^-}.
\end{equation}
Working in the eigenbasis of the qubit, we have:
\begin{equation}
    H_n(\rho_q(t)) = \frac{1}{1-n}\log\left(\sum_i\lambda_i^n\right),\\[1em]
\end{equation}
\begin{equation}
    \dot{H}_n(\rho_q(t)) = \frac{\sum\limits_i n\lambda_i^{n-1}\dot{\lambda}_i}{\sum\limits_i \lambda_i^n},
\end{equation}
\begin{equation}
    \ddot{H}_n(\rho_q(t)) = \frac{1}{1-n}\left(\frac{\sum\limits_i\big(n(n-1)\lambda_i^{n-2}\dot{\lambda}_i^2 + n\lambda_i^{n-1}\ddot{\lambda}_i\big)}{\sum\limits_i \lambda_i^n} - \frac{\Big(\sum\limits_i n\lambda_i^{n-1}\dot{\lambda}_i\Big)^2}{\Big(\sum\limits_i\lambda_i^n\Big)^2}\right).
\end{equation}
Let us therefore consider the eigenvalues of $\rho_q(t)$. Diagonalising Equation \ref{eq:time_ev_rho}, we obtain:
\begin{equation}
    \lambda^{\pm}(t) = \frac{1\pm\sqrt{1-4(\delta - \delta^2 - \vert\alpha\vert^2 e^{-4t^2})}}{2},
\end{equation}
such that 
\begin{equation}
    \dot{\lambda}^{\pm}(t) = \mp\frac{8\vert\alpha\vert^2 t e^{-4t^2}}{\sqrt{1-4(\delta - \delta^2 - \vert\alpha\vert^2 e^{-4t^2})}},
\end{equation}
and 
\begin{equation}
    \ddot{\lambda}^{\pm}(t) = \mp\left(\frac{8\vert\alpha\vert^2e^{-4t^2}\big(1 -8t^2\big)}{\sqrt{1-4(\delta - \delta^2 - \vert\alpha\vert^2 e^{-4t^2})}}+\frac{128\vert\alpha\vert^4 t^2 e^{-8t^2}}{\big(1-4(\delta - \delta^2 - \vert\alpha\vert^2 e^{-4t^2})\big)^{3/2}}\right).
\end{equation}
From here, we readily see that $\dot{\lambda}^{\pm}\vert_{t=0} = 0$, such that
\begin{equation}
    \ddot{H}_n(\rho_q(t=0)) = \frac{n\sum\limits_i \lambda_i^{n-1}\ddot{\lambda}_i}{(1-n)\sum\limits_i\lambda_i^n}\Bigg\vert_{t=0},
\end{equation}
where
\begin{equation}
    \ddot{\lambda}^{\pm}\big\vert_{t=0} = \mp \frac{8\vert\alpha\vert^2}{\sqrt{1-4(\delta - \delta^2 - \vert\alpha\vert^2)}}
\end{equation}
Denoting the $t=0$ eigenvalues  as $\lambda^{\pm}$, we finally obtain:
\begin{equation}\label{eq:qubit_result}
    \ddot{H}_n(\rho_q(t=0)) = \frac{-8n\vert\alpha\vert^2\big({\lambda^{-}}^{n-1} - {\lambda^{+}}^{n-1}\big)}{\big(n-1\big)\big(\lambda^+ - \lambda^-\big)\big({\lambda^{-}}^n + {\lambda^{+}}^n\big)}
\end{equation}
Let us now compare this result to Equation \ref{eq:deriv_gen}. In Equation \ref{eq:deriv_gen}, all quantities are evaluated at $t=0$. Therefore, we simply work with the initial qubit state $\rho_q$. In order to then evaluate Equation \ref{eq:deriv_gen}, we re-express the system in the eigenbasis of $\rho_q$. We have:
\begin{equation}
    \lambda^{\pm} = \frac{1\pm\sqrt{1-4(\delta -\delta^2-\vert\alpha\vert^2)}}{2},
\end{equation}
\begin{equation}
    \ket{\lambda^+} = \frac{1}{\sqrt{1+\vert x\vert^2}}
    \Big(\ket{z^+} + x\ket{z^-}\Big), \qquad 
        \ket{\lambda^-} = \frac{1}{\sqrt{1+\vert y\vert^2}}
    \Big(\ket{z^+} + y\ket{z^-}\Big),
\end{equation}
where
\begin{equation}
    x := \frac{-(\delta - \lambda^+)}{\alpha}, \qquad y := \frac{1-(\delta + \lambda^+)}{\alpha}.
\end{equation}
We may now re-express $\sigma_z = \ket{z^+}\bra{z^+} - \ket{z^-}\bra{z^-}$ in terms of the eigenbasis of $\rho_q$. We have:
\begin{align}
    \sigma_z = \frac{1}{\vert x-y\vert^2}\bigg(& (1+\vert x\vert^2)(\vert y\vert^2 -1)\ket{\lambda^+}\bra{\lambda^+}\nonumber\\
    &+\sqrt{(1+\vert x\vert^2)(1+\vert y\vert^2)}(1-yx^*)\ket{\lambda^+}\bra{\lambda^-}\nonumber\\
    &+\sqrt{(1+\vert x\vert^2)(1+\vert y\vert^2)}(1-xy^*)\ket{\lambda^-}\bra{\lambda^+}\nonumber\\    
    &+(1+\vert y\vert^2)(\vert x\vert^2 -1))\ket{\lambda^-}\bra{\lambda^-}\bigg).
\end{align}
Working in this eigenbasis, we have:
\begin{align}
    \Tr_q\Big[\rho_q^{n-1}[\sigma_z,\rho_q]\sigma_z\Big] &= \sum_{i,j}\lambda_j^{n-1}(\lambda_i-\lambda_j)\vert{\sigma_z}_{ij}\vert^2\nonumber\\
    &=\big({\lambda^-}^{n-1}-{\lambda^+}^{n-1}\big)\big(\lambda^+ - \lambda^-\big)\vert {\sigma_z}_{+-}\vert^2\nonumber\\
    &=\frac{4\vert\alpha\vert^2\big({\lambda^-}^{n-1}-{\lambda^+}^{n-1}\big)}{\big(\lambda^+ - \lambda^-\big)},
\end{align}
where the characteristic equation, ${\lambda^+}^2 - \lambda^+ + \delta - \delta ^2 - \vert\alpha\vert^2 =0$, is used to convert the $\sigma_z$ components into expressions in terms of $\lambda^{\pm}$. Taking into account that the field variance in this case is $1$, and substituting the above result into Equation \ref{eq:deriv_gen}, we again arrive at Equation \ref{eq:qubit_result} for $\ddot{H}_n(\rho_q(t=0))$, which indeed confirms the validity of the perturbative approach we have employed in the previous sections. 

Using these non-perturbatively verified results, we can assess how the $n$-exposure depends upon the precise configuration of the qubit. We have:
\begin{equation}
    E_{n,q} = (\Delta \sigma_z)^2 - D_{n, q} = 4(\delta - \delta^2) +\frac{4\vert\alpha\vert^2\left({\lambda^-}^{n-1} - {\lambda^+}^{n-1}\right)}{\left(\lambda^+-\lambda^-\right)\left({\lambda^-}^{n} + {\lambda^+}^{n}\right)}.
\end{equation}
From this expression, we see that the $n$-exposure does not depend on the phase of $\alpha$. Indeed, the $n$-exposure can be expressed in terms of $\delta$ and $\vert\alpha\vert^2$ alone. 

\begin{figure}[!htb]
\centering
\begin{minipage}{0.4\textwidth}
    \centering
    \includegraphics[scale = 0.55, trim={4cm 0cm 0cm 0cm}]{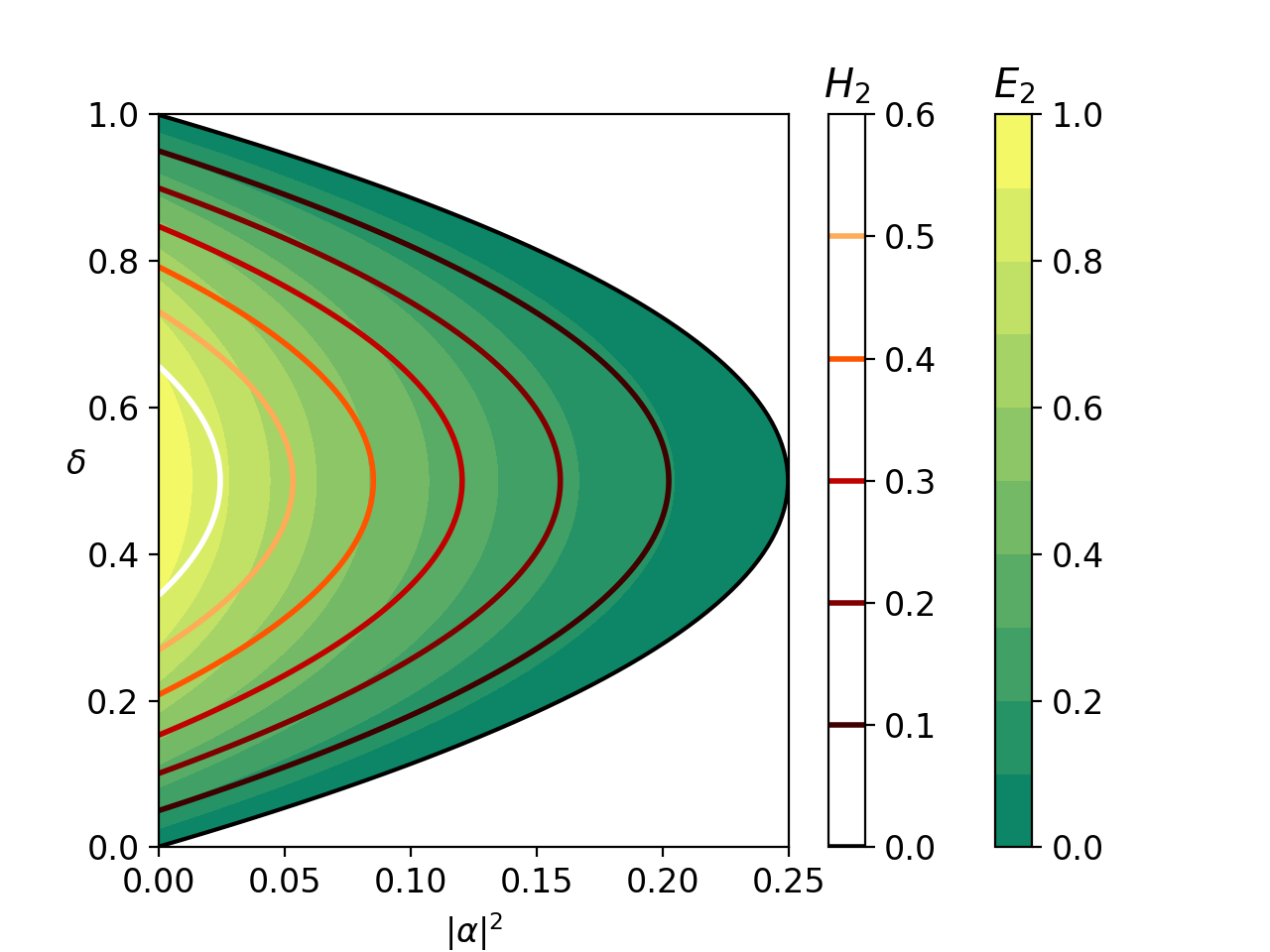}
\end{minipage}
\begin{minipage}{0.4\textwidth}
    \centering
    \includegraphics[scale = 0.55, trim={0cm 0cm 0cm 0cm}]{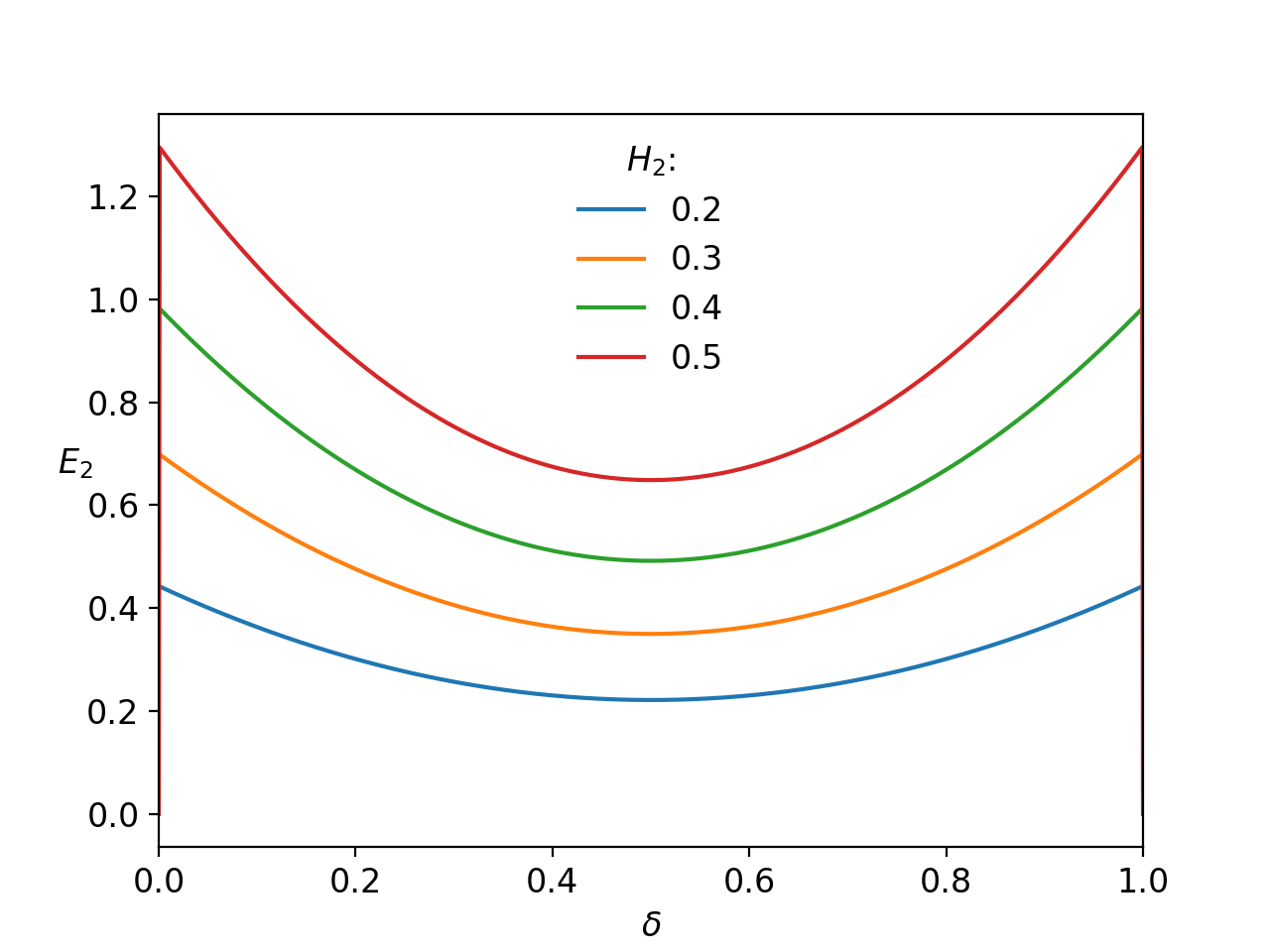}
\end{minipage}
\caption{Left: contour plot of the 2-exposure of the qubit across the $(\delta,\vert\alpha\vert^2)$ plane (green). Also included are contours of the 2-R\'enyi entropy, which do not align with the 2-exposure contours. Right: variation in the 2-exposure as a function of $\delta$ for constant values of 2-R\'enyi entropy.}\label{fig:non-bloch}
\end{figure}

As discussed in Section \ref{sec:n_1}, the leading order behaviour of the $n\rightarrow 1$ coherent information is non-trivial. This is due to the divergence in the second time derivative of the von Neumann entropy of the environment system.\footnote{Note that the divergent second derivative does not imply a divergence in the von Neumann entropy itself - see Appendix \ref{app:exact_plots}.} Hence, we will restrict our attention to $n>1$ and choose the simplest case ($n=2$) to illustrate the features of the $n$-exposure. In Figure \ref{fig:non-bloch}, we present a contour plot of the 2-exposure as a function of these two variables.\footnote{Note that the positivity of the eigenvalues requires $\vert\alpha\vert^2 \leq \delta - \delta ^2$.} We also include isocurves of the 2-R\'enyi entropy. Crucially, we see that the isocurves of the exposure follow different trajectories to the isocurves of the entropy through the $(\delta,\vert\alpha\vert^2)$ plane. This illustrates that the leading order change in the 2-coherent information is not simply a function of the total amount of entanglement present, but depends upon the precise configuration of the qubit. The right hand side of Figure \ref{fig:non-bloch} highlights this phenomenon by plotting the variation in the 2-exposure along isocurves of constant 2-R\'enyi entropy.

\begin{figure}[!htb]
\centering
\includegraphics[scale = 0.6, trim={0cm 2cm 0cm 0.5cm}]{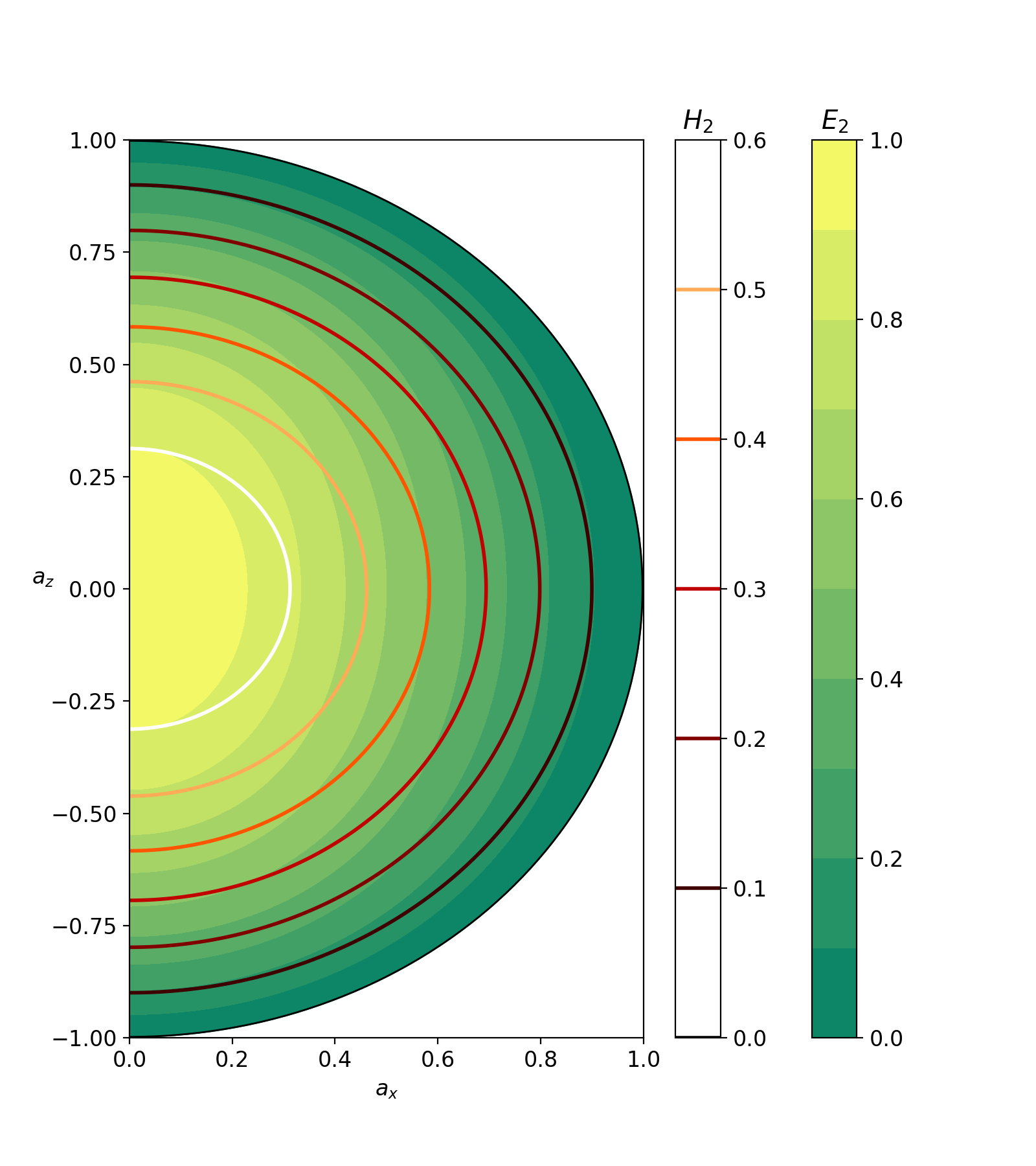}
\caption{2-exposure of the qubit in the Bloch sphere representation. Again we see that the isocurves of the 2-R\'enyi entropy are disaligned with those of the exposure, indicating that it is not simply the amount of entanglement, but the way in which it is distributed that determines the exposure. In this case, if avoidance of transmission is desired then the equatorial direction is `safer' than the polar direction.}\label{fig:bloch}
\end{figure}

One may also consider the $n$-exposure of the qubit in the Bloch sphere representation. We first set:
\begin{equation}
    \rho_q = \frac{1}{2}(\mathbb{I} + a_x\sigma_x +a_y\sigma_y +a_z\sigma_z),
\end{equation}
where the Bloch vector is $\Vec{a}=(a_x,a_y,a_z)$. This then leads to the following transformation:
\begin{equation}
    \delta = \frac{1+a_z}{2}, \qquad \vert\alpha\vert^2 = \frac{\vert a_x\vert^2 + \vert a_y\vert^2}{4}.
\end{equation}
We may use these expressions to convert Figure \ref{fig:non-bloch} into the Bloch sphere representation. We note that in this representation, the exposure is independent of the phase in the $(a_x, a_y)$ plane, so that it is sufficient to plot only a cross section in the $(a_x, a_z)$ plane. We illustrate the 2-exposure in the Bloch sphere representation in Figure \ref{fig:bloch}. Again, we also plot isocurves of the 2-R\'enyi entropy to illustrate that the exposure is a non-trivial function of the distribution of the entanglement in the state space of the qubit. 

In particular, we note that in order to minimise exposure at the onset of an interaction, we should tune the distribution of the entanglement in the initial system. In this example we see from Figure \ref{fig:bloch} that for the same initial 2-R\'enyi entropy, the 2-exposure is minimised along the $a_x$ axis, while it is maximised along the $a_z$ axis. This indicates the optimal state of the qubit in the Bloch sphere representation in order to minimise the leading order change in the second R\'enyi entropy. The same analysis can be applied to more complex systems in order to identify the ideal initial configuration for the minimisation or maximisation of the exposure.

Figures \ref{fig:non-bloch} and \ref{fig:bloch} illustrate that in the qubit case, the 2-exposure is always positive. This is due to the simplicity of the qubit state space. In the next section, however, we will consider a qutrit system, for which the higher dimensionality leads to more complex results.

\section{Application to qutrit systems}\label{sec:qutrit}

In this section, we provide a visualisation of the 2-exposure for a simple qutrit system. As we augment the qutrit density matrix, we will see how the exposure can vary in non-trivial ways, and in particular, that it can change sign. 
To this end, let us first consider the general form of the qutrit density matrix:

\begin{equation}
\rho = 
    \begin{pmatrix}
        \omega_{x} & \frac{-ia_z - q_z}{2} & \frac{-ia_y - q_y}{2} \\
        \frac{ia_z - q_z}{2} & \omega_{y} & \frac{-ia_x - q_x}{2} \\
        \frac{ia_y - q_y}{2} & \frac{ia_x - q_x}{2} & \omega_{z} 
    \end{pmatrix}.
\end{equation}
While this generic qutrit state possesses eight degrees of freedom, we may restrict our attention to the case where all off-diagonal matrices are purely imaginary ($q_j = 0$). We then have:
\begin{equation}
\rho = 
    \begin{pmatrix}
        \omega_{x} & \frac{-ia_z}{2} & \frac{-ia_y}{2} \\
        \frac{ia_z}{2} & \omega_{y} & \frac{-ia_x}{2} \\
        \frac{ia_y}{2} & \frac{ia_x}{2} & \omega_{z} 
    \end{pmatrix},\\[1em]
\end{equation}
where $0\leq a_j \leq 1$, $0\leq \omega_j \leq 1$, and $\sum_j \omega_j = 1$. The non-negativity of this simplified density matrix is ensured through the following conditions \cite{Kurzynski2016}:

\begin{equation}\label{eq:cond_1}
    4\omega_j\omega_k \geq a_l^2,
\end{equation}

\begin{equation}\label{eq:cond_2}
    4\omega_j\omega_k\omega_l \geq \omega_j a_j^2 + \omega_k a_k^2 + \omega_l a_l^2.
\end{equation}
From here, we can simplify further by enforcing $\omega_j = \omega_k = \omega_l = 1/3$. In this case, conditions \ref{eq:cond_1} and \ref{eq:cond_2} can be re-written as a single condition, namely

\begin{equation}
    \frac{4}{9} \geq a_x^2 + a_y^2 + a_z^2,
\end{equation}
which describes a space of $a_x, a_y$, and $a_z$ values enclosed by an octant of a sphere of radius $2/3$. Given a choice of an operator acting on the qutrit system, we may then examine the variation of the exposure as we move through this space of states.

The unitary dynamics of a qutrit can be described through a set of three different types of transformations \cite{Kurzynski2016}. Namely, rotations, single-axis twisting, and dual-axes counter-twisting. We can build the full space of Hamiltonians from the following three matrices:

\begin{equation}
    S_x = 
    \begin{pmatrix}
        0 & 0 & 0 \\
        0 & 0 & -i \\
        0 & i & 0 
    \end{pmatrix}, \qquad
    S_y =
        \begin{pmatrix}
        0 & 0 & i \\
        0 & 0 & 0 \\
        -i & 0 & 0 
        \end{pmatrix}, \qquad
    S_z =
        \begin{pmatrix}
        0 & -i & 0 \\
        i & 0 & 0 \\
        0 & 0 & 0 
        \end{pmatrix}.
\end{equation}

\begin{figure}[!htb]\label{fig:qutrits}
\centering
\begin{minipage}{0.4\textwidth}
    \centering
    \includegraphics[scale = 0.55, trim={4cm 0cm 0cm 0cm}]{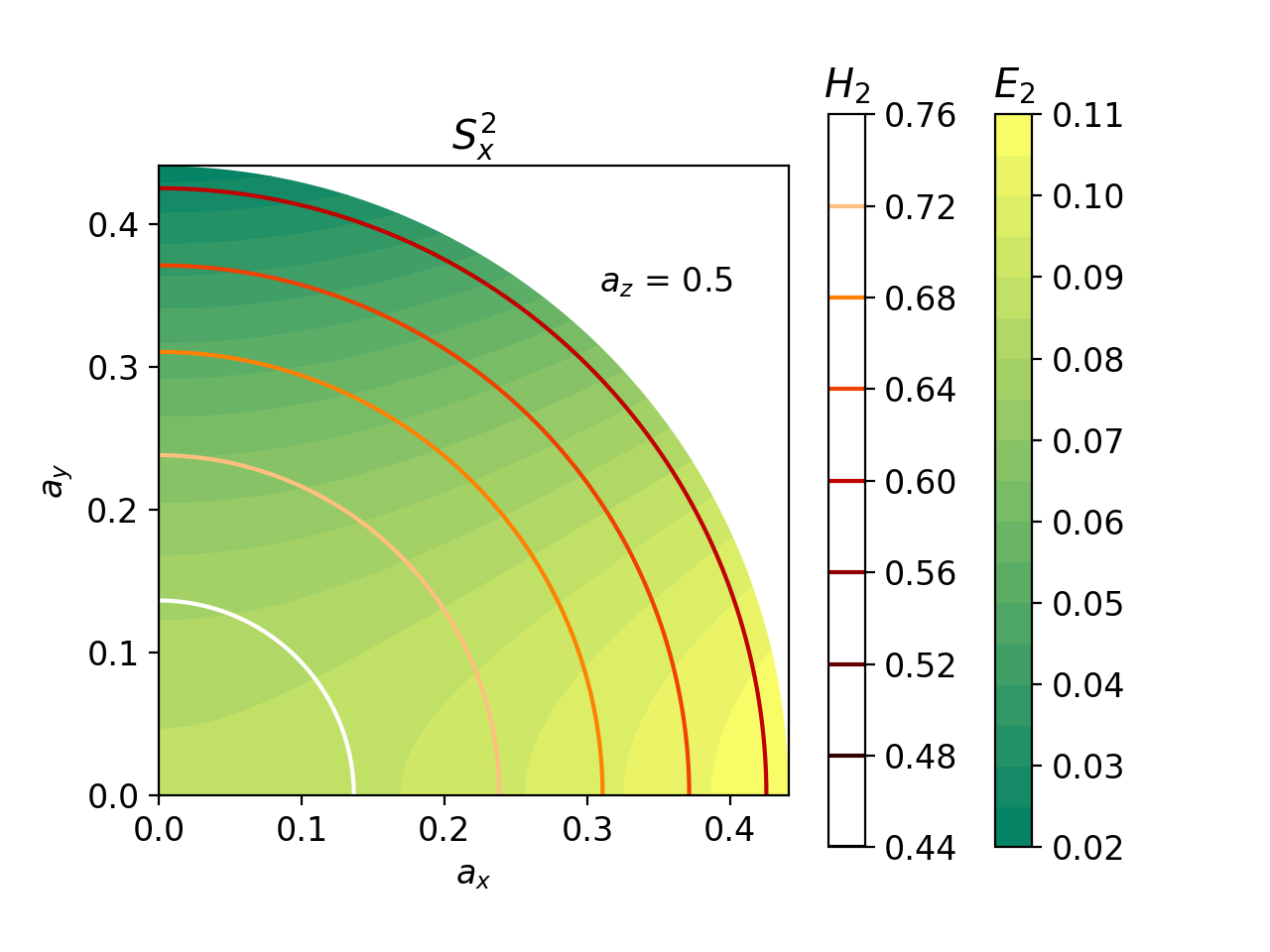}
\end{minipage}
\begin{minipage}{0.4\textwidth}
    \centering
    \includegraphics[scale = 0.55, trim={0cm 0cm 0cm 0cm}]{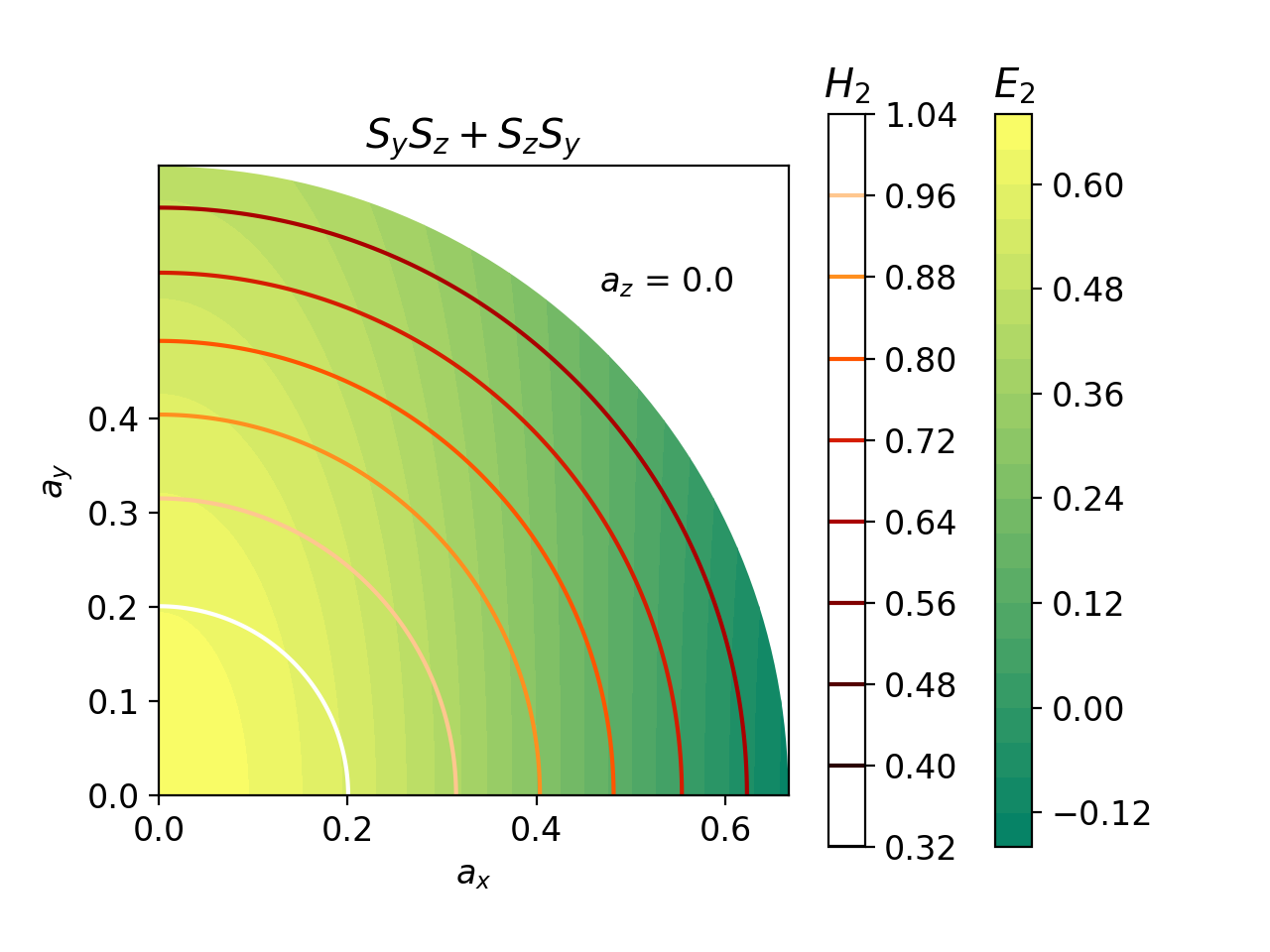}
\end{minipage}
\caption{Left: 2-exposure across the $a_x,a_y$ plane at $a_z=0.5$ for the Hamiltonian $S_x^2$. Right: 2-exposure for the Hamiltonian $S_yS_z+S_zS_y$ across the $a_x,a_y$ plane at $a_z=0.0$. In this case we see regions of negative exposure. Also plotted are the 2-R\'enyi entropy isocurves.}
\end{figure}

We may now calculate the 2-exposure for different qutrit configurations and different Hamiltonians according to Equation \ref{eq:ee}. Since our space of possible states is a three-dimensional octant of a sphere, we represent the results as cross-sections in the $(a_x, a_y)$ plane. In Figure \ref{fig:qutrits} we illustrate the exposure at the $a_z=0.5$ plane for two arbitrary Hamiltonians. We also include isocurves of constant 2-R\'enyi entropy, demonstrating that the $n$-exposure is not simply dependent upon the total amount of entanglement present, but rather on the precise distribution of this entanglement. For each example Hamiltonian we also see that, depending upon the precise configuration of the input state, the 2-exposure can be either positive or negative, indicating, e.g., in the case where $B$ is an environment,  `safe' and `unsafe' configurations, respectively.

\section{Conclusions and Outlook}\label{sec:conclusion}

In this work, we have considered a setup where a system $A$ is initially entangled with and purified by an ancilla $\tilde{A}$. System $A$ then starts to interact with a system $B$ and may, therefore, become entangled with $B$. Through the interaction, $A$ may lose some of its quantum correlations with $\tilde{A}$ and $B$ may acquire quantum correlations with $\tilde{A}$. We identified the maps of reduced density matrices, from $A$ to $A'$ and from $A$ to $B'$ (primes denoting a later time) the direct and the complementary channel respectively. 

Because the quantum capacity of a quantum channel is defined by optimization of the coherent information over input states and parallel channel uses, we have calculated the leading order change in coherent information of each channel. This, in turn, poses a lower bound on the quantum channel capacity. 

In quantum technologies, depending on the application, it can be desirable to
maximize the direct channel, in order to protect existing quantum correlations from leakage to an environment $B$. Conversely, it may be desirable to maximize the complementary channel for purposes of quantum communication from $A$ to $B$. It is therefore of interest to determine the factors which govern the leading order change in coherent information at the onset of a quantum channel, so that quantum systems may be initialised in a configuration most favourable to the application.

To calculate the leading order behaviour of the coherent information, we first showed that the free Hamiltonians do not contribute to leading (second) order. This dramatically simplifies the calculations, as the interaction Hamiltonian may be considered alone. Moreover, this has the important implication that resonance phenomena, which require free Hamiltonians, do not contribute to leading order. This is of particular significance in systems which make use of the light-matter interaction, in which resonance phenomena are commonly encountered. The significance of this finding will be investigated further in future work.

We then found that the leading order term in the perturbative expansion of the coherent information diverges when system $B$ is initially pure. We therefore generalized the notion of coherent information to the family of $n$-coherent information, via the $n$-R\'enyi entropies. The $n$-coherent information is always regular for $n>1$. 
Calculation of the dynamics of the $n$-coherent information, and therefore of the $n$-R\'enyi entropies, is desirable  because it implicitly also determines the dynamics of the entire spectrum of the underlying density matrix. This is because the family of $n$-R\'enyi entropies (unlike the von Neumann entropy alone) is sufficient to reconstruct the spectrum of the density matrix.  

A further, useful property of the $n$-coherent information arises from the fact that  the $n$-R\'enyi entropies of two subsystems that together constitute a pure supersystem are identical, as is the case also for the von Neumann entropy. Therefore, and because $H_n(\tilde{A})^{''}=0$, calculating the leading order $n$-R\'enyi entropies and $n$-coherent informations of the direct and complementary channels implies also knowledge of the leading order of all other coherent informations that can be formed among the three subsystems $A,\tilde{A},B$. For example, when $(H_n(A)-H_n(AB))$ becomes positive, this indicates that quantum correlations between $A$ and $B$ are becoming established, and we have  $(H_n(A)-H_n(AB))^{''}=H_n(A)^{''}$ since $H_n(AB)^{''} = H_n(\tilde{A})^{''}=0$. 

As it is known that the traditional (1-)coherent information serves as a lower bound on the quantum channel capacity, it is interesting to ask whether the $n-$coherent information performs an analogous role within quantum information. Indeed, through weak subadditivity, we find that the $n-$coherent information plays an important role in determining bounds on quantum communication complexity \cite{2002quant.ph..4093V}. Whereas the quantum channel capacity represents the highest rate at which quantum information may be transferred over a noisy quantum channel, the quantum communication complexity represents the minimum number of qubits which must be transferred in order to perform a particular quantum transformation. In future work, we will explore the connection between these bounds on quantum communication in greater detail.

We then explicitly calculated the leading order time evolution of the $n-$coherent information of the direct and complementary channels, for the case of interaction Hamiltonians of the form $\hat{A}\otimes \hat{B}$ and showed that the leading order is determined by a quantity that we call the $n$-exposure. Intuitively, the $n$-exposure quantifies how much of the pre-existing quantum correlations in system $A\tilde{A}$ are exposed to the interaction Hamiltonian between systems $A$ and $B$. The larger the exposure, the faster does the coherent information of the direct and complementary channel change. Importantly, we have shown that the $n-$exposure is not proportional to the amount of pre-existing entanglement (as measured by the von Neumann entropy), nor is it proportional to the $n-$R\'enyi entropy. Instead, it depends non-trivially on the initial state of $A$, such that two initial configurations with the same initial entanglement may exhibit entirely different behaviours at the onset of the quantum channel. We illustrated how the $n$-exposure dependends on the initial state of $A$, and on the interaction Hamiltonian, in a simple model of the light-matter interaction and in qutrit systems. 

In future work, it will be very interesting to calculate the analog of the $n$-exposure for interaction Hamiltonians of the more general form $\sum_r \hat{A}_r\otimes\hat{B}_r$. Interestingly, in this case the contributions to the leading order change in the $n$-coherent information of the direct and complementary channels do not simply factorize into separate contributions from systems $A$ and $B$. It will then also be interesting to revisit the limit $n\rightarrow 1$ (see also Appendix \ref{app:caution}).

In practice, for example in the running of quantum processors, it may be possible to use the notion of exposure to optimize their performance, for example, by extremizing the exposure to maximize a desired transfer of quantum correlations, or conversely, to minimize exposure in order to reduce leakage of quantum correlations to an environment. 
We also anticipate that the notion of $n-$exposure may prove useful in the implementation of quantum algorithms when certain qubits are idle. In particular, one may rotate a qubit in its Bloch sphere along an isocurve of $n$-R\'enyi entropy, thus preserving its $n$-coherent information with the rest of the qubits, but to a location in the Bloch sphere that decreases its exposure to decohering. This principle should also be extendable to multiple qubits at a time, possibly utilizing supervised machine learning with a cost function that contains the exposure. Work in this direction is in progress.  

In this work we have focused on the dynamics of the coherent information, 
which we note is not itself an entanglement monotone. Indeed, given that our study involves a tripartite system, traditional measures of entanglement such as the von Neumann entropy are insufficient to describe true tripartite entanglement. Hence, work is ongoing to extend our results using the logarithmic negativity as an entanglement monotone of tripartite systems. Nevertheless, positivity of the coherent information does imply the presence of quantum correlations, and coherent information remains a critical tool in the assessment of quantum channel capacity. 

$$$$

\bf Acknowledgements. \rm AK acknowledges support through a Discovery Grant of the National Science and Engineering Council of Canada (NSERC), a Discovery Project grant of the Australian Research Council (ARC), and a Google Faculty Research Award. 
B\v{S} is supported in part by the Perimeter Institute, which is supported in part by the Government of Canada through the Department of Innovation, Science
and Economic Development Canada and by the Province of Ontario through the Ministry of Economic Development, Job Creation and
Trade.

\section*{References}
\bibliographystyle{iopart-num}  
\bibliography{refs}

\appendices

\section{Non-perturbative calculation}\label{app:non-pert}

While we have taken a perturbative approach throughout this work, we illustrate here the exact calculation of the time-evolved state of System $A$. We find that the exact form of $\rho_A$ is not easily raised to the $n^\text{th}$ power in general, as is required for the calculation of the $n$-R\'enyi entropies. While there are particularly simple interaction Hamiltonians such as that invoked in Section \ref{sec:rabi} which enable exact calculations to be performed, this is not often the case. 
This highlights the utility of our perturbative method. Let us first express the time-evolved state of system $A$ as:
\begin{align}
    \rho_{A}(t) &= \Tr_{B}\Big[e^{it\hat{A}\otimes\hat{B}}\rho_A\otimes\rho_B e^{-it\hat{A}\otimes\hat{B}}\Big]\nonumber\\[1em]
    &= \sum_r \bra{b_r}e^{it\hat{A}\otimes\hat{B}}\rho_A\otimes\rho_B e^{-it\hat{A}\otimes\hat{B}}\ket{b_r}\nonumber\\[1em]
    &= \sum_r e^{it\hat{A}b_r}\rho_Ae^{-it\hat{A}b_r}\rho_{Brr},
\end{align}
where, working in the eigenbasis of $\hat{B}$:
\begin{equation}
    \rho_{Bij} = \bra{b_i}\rho_B\ket{b_j}.
\end{equation}
Representing $\rho_A$ in the eigenbasis of $\hat{A}$:
\begin{equation}
    \rho_A = \sum_{ij}\rho_{Aij}\ket{a_i}\bra{a_j},
\end{equation}
we have:
\begin{equation}
    \rho_{A}(t) = \sum_{ijk}\rho_{Aij}\rho_{Bkk}e^{itb_k(a_i-a_j)}\ket{a_i}\bra{a_j},
\end{equation}
Such that
\begin{equation}
    \gamma_{A,n}(t)=\Tr_A[\rho_A(t)^n],
\end{equation}
where $\rho_A(t)$ is a matrix whose matrix element $\rho_A(t)_{ij}$ is given by
\begin{equation}
    \rho_A(t)_{ij} = \rho_{Aij}\sum_k e^{itb_k(a_i-a_j)}\rho_{Bkk}.
\end{equation}

\section{Qubit von Neumann entropy in quantum Rabi model}\label{app:exact_plots}

\begin{figure}[!htb]
\centering
\begin{minipage}{0.4\textwidth}
    \centering
    \includegraphics[scale = 0.55, trim={4cm 0cm 0cm 0cm}]{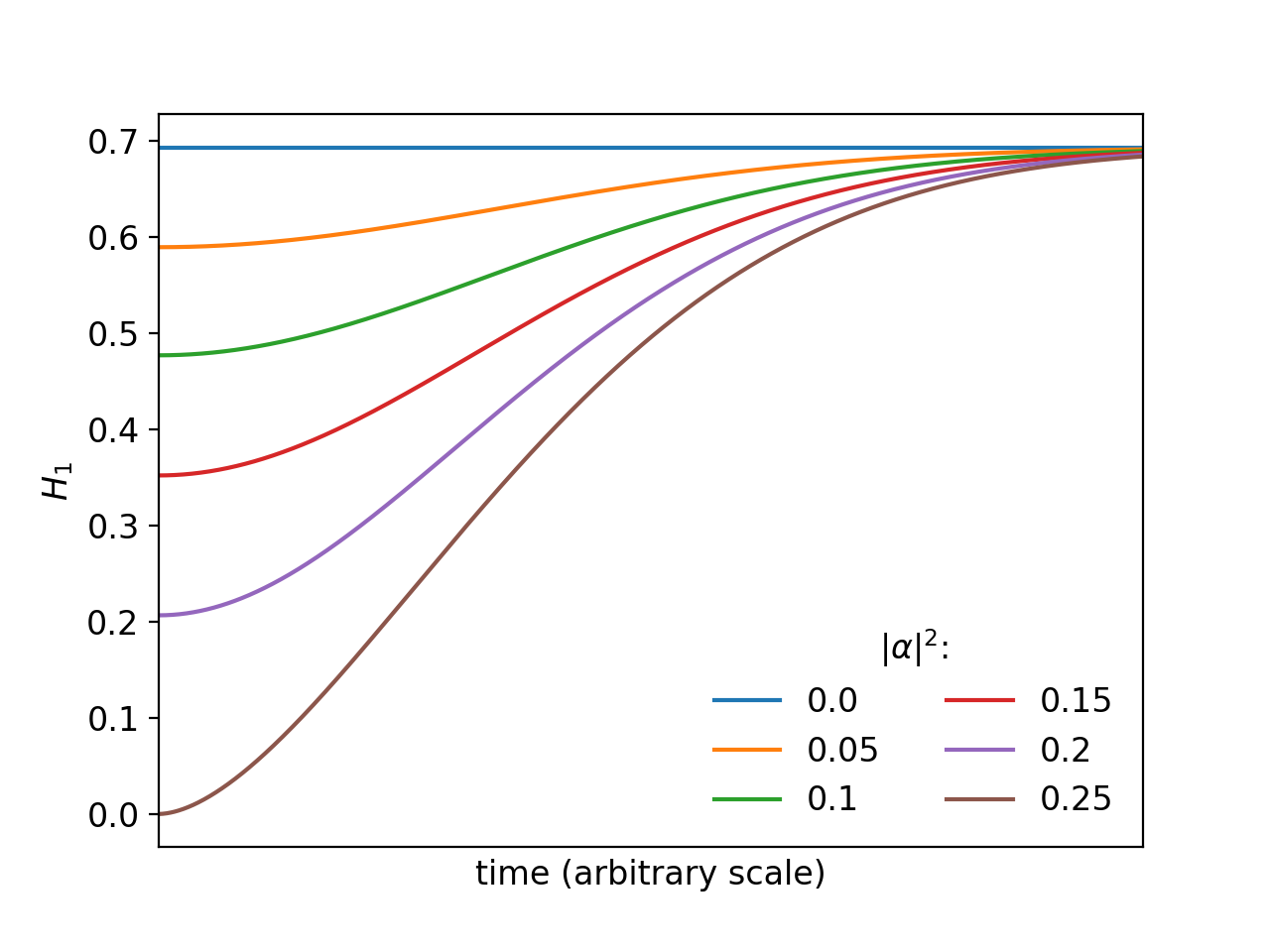}
\end{minipage}
\begin{minipage}{0.4\textwidth}
    \centering
    \includegraphics[scale = 0.55, trim={0cm 0cm 0cm 0cm}]{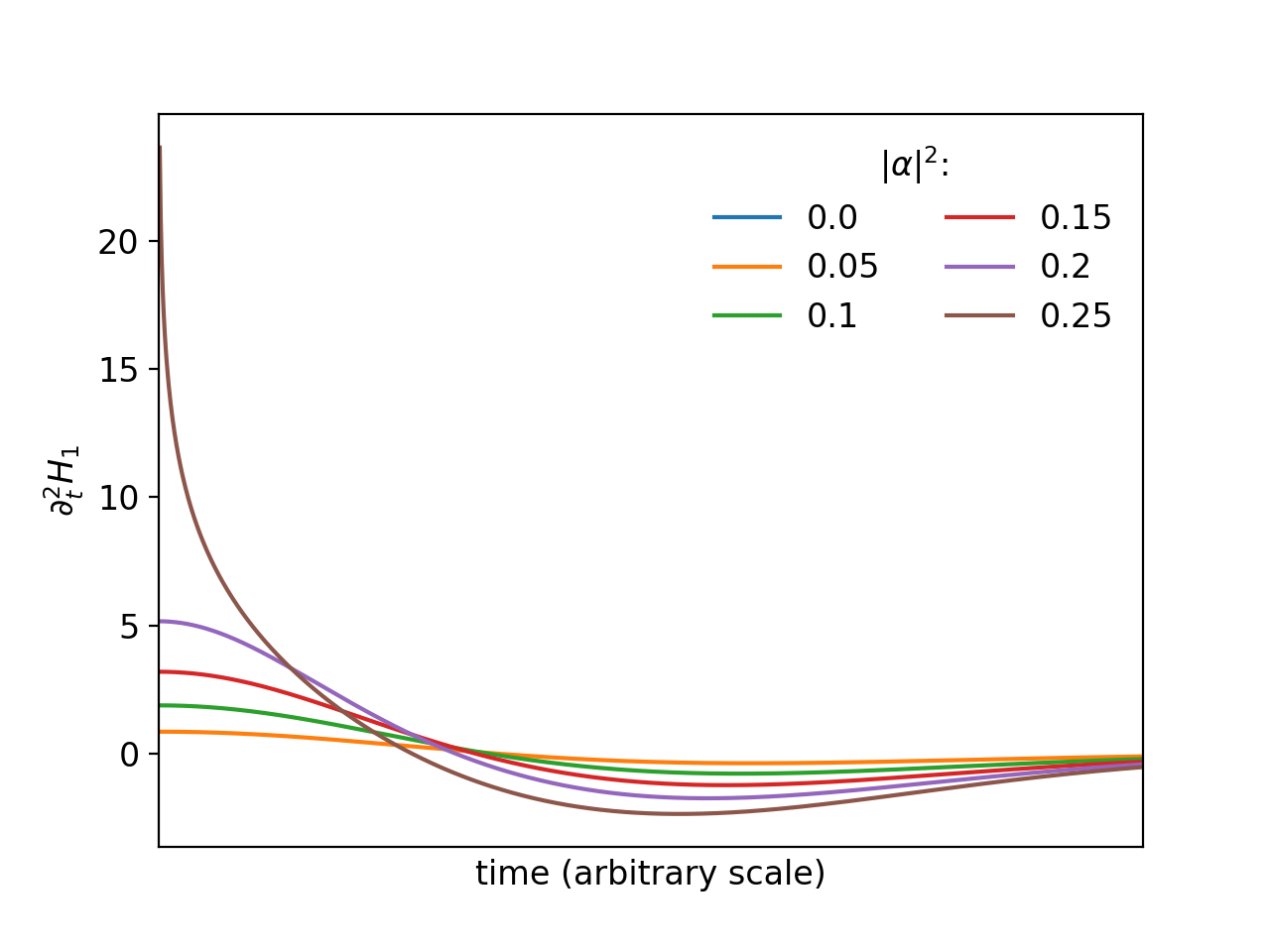}
\end{minipage}
\caption{Left: the exact time evolution of the von Neumann entropy of the qubit under the quantum Rabi interaction. The $\vert\alpha\vert^2 = 0.25$ case corresponds to a pure input state. Right: the second time derivative of the von Neumann entropy of the qubit for the quantum Rabi interaction. While the second derivative is infinite at $t=0$ for a pure state, this quickly falls off to a finite (indeed negative) value.}\label{fig:time_ev}.  
\end{figure}

In fig. \ref{fig:time_ev}. we plot the exact time evolution of the von Neumann entropy of a qubit system undergoing the quantum Rabi interaction as described in Section \ref{sec:rabi}. We illustrate that, while the second time derivative diverges for a pure input state, the von Neumann entropy itself remains finite. Nevertheless, the divergence of the second time derivative renders the second order perturbative analysis of the (1-)coherent information unsuitable as a measure of entropy transfer.

\section{Numerical illustration of the noncommutativity of the limits $\epsilon \rightarrow 0$ and $\lambda \rightarrow 0$.}
\label{numer}

\begin{figure}[!htb]
\centering
\begin{minipage}{0.4\textwidth}
    \centering
    \includegraphics[scale = 0.55, trim={4cm 0cm 0cm 0cm}]{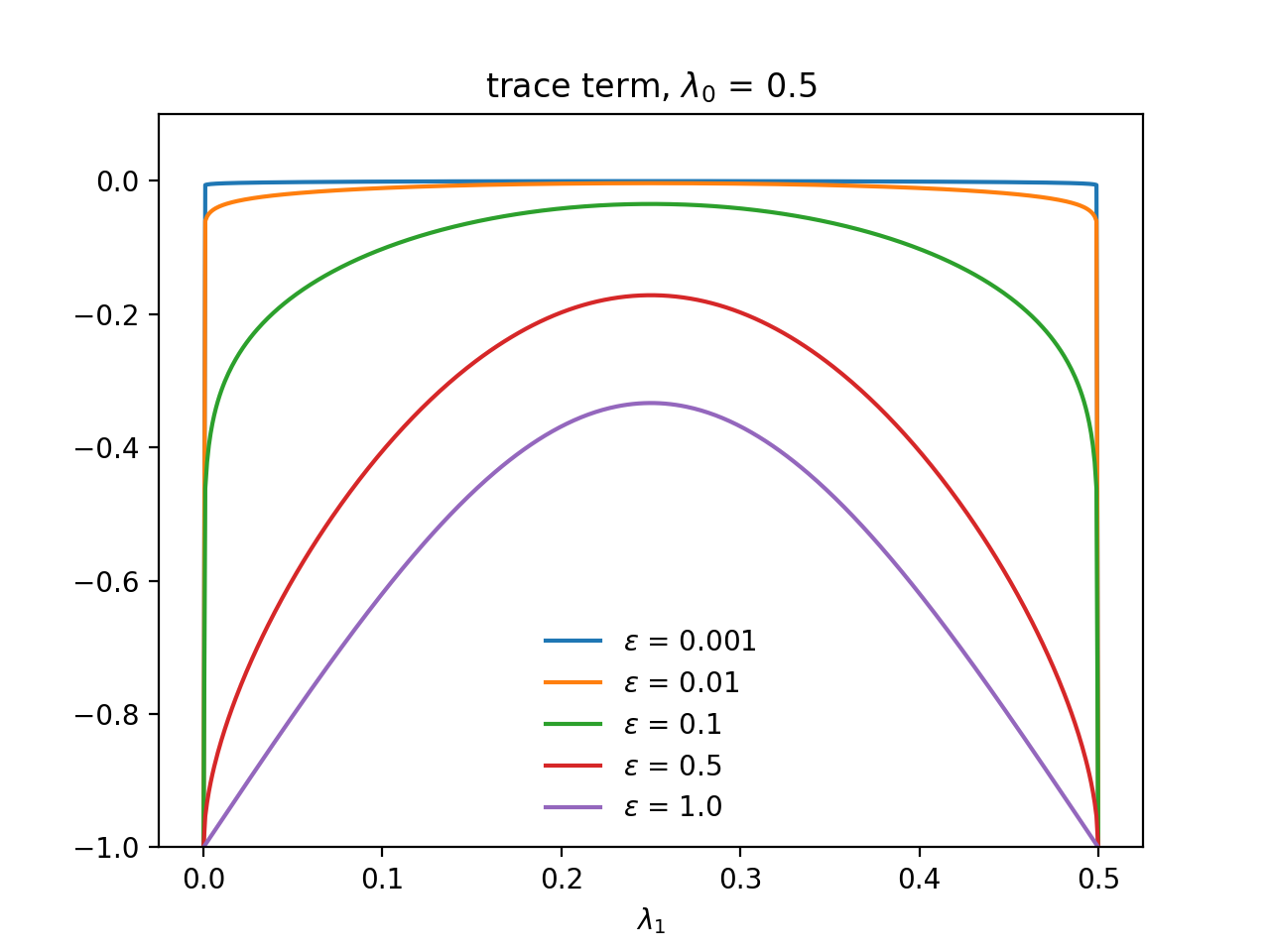}
\end{minipage}
\begin{minipage}{0.4\textwidth}
    \centering
    \includegraphics[scale = 0.55, trim={0cm 0cm 0cm 0cm}]{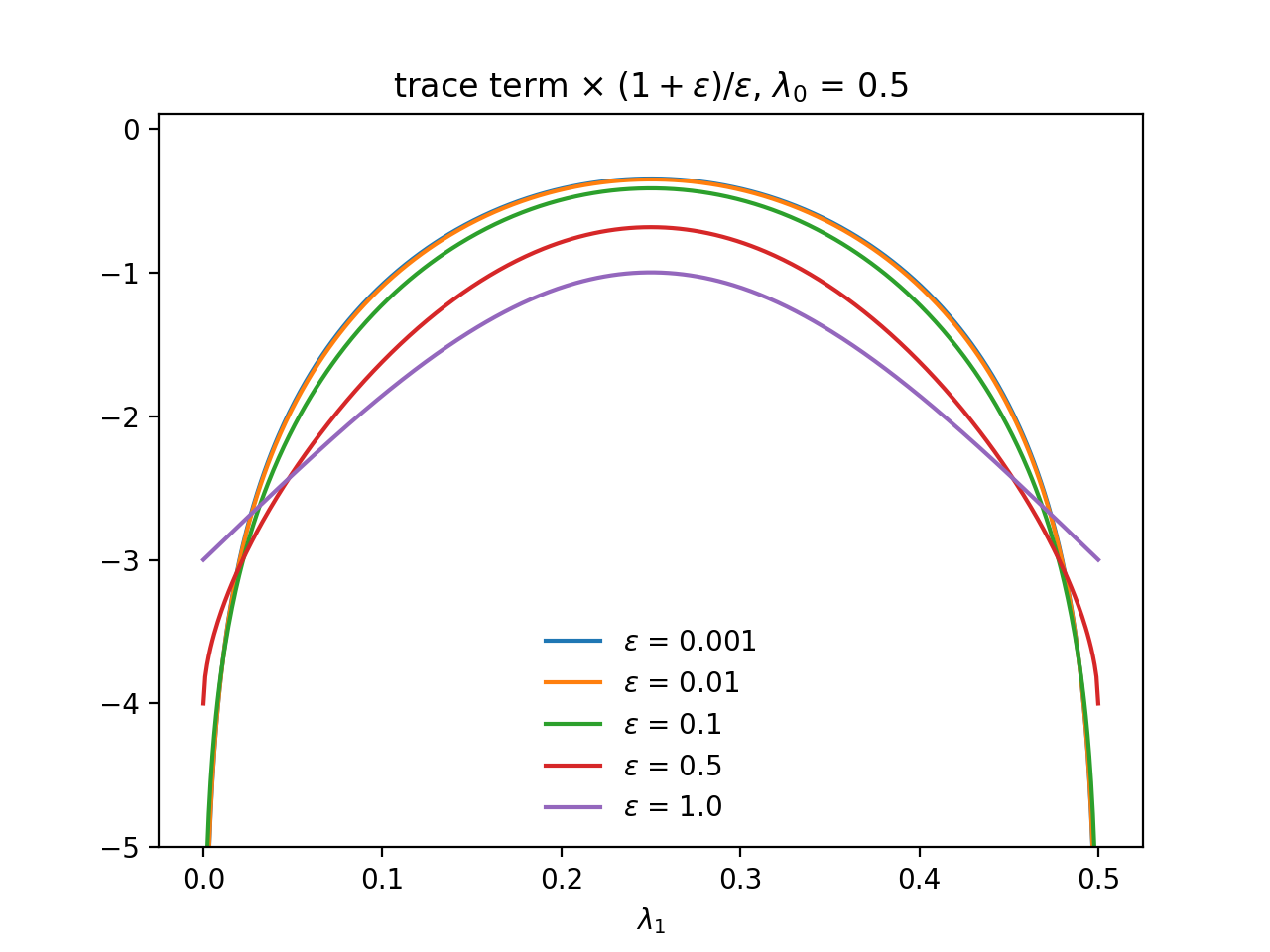}
\end{minipage}
\caption{Left: plot of the sum \ref{eq:to_plot_num_0}, where we see that this tends to zero for small $\varepsilon$ except where an eigenvalue approaches zero, at which point an abrupt jump is observed. 
Right: plot of Equation \ref{eq:to_plot_num}. Here we see that the jump in the left-hand plot is smoothed, but as one eigenvalue approaches zero the expression approaches a divergence.}\label{fig:n_1_numerical}.  
\end{figure}

We expect that for finite eigenvalues in the small $\varepsilon$ limit the trace term in Equation (\ref{trte}) reduces to:
\begin{equation}\label{eq:zero_lim}
    \lim_{\varepsilon\rightarrow 0}\sum_{i,j} (\lambda_j^{\varepsilon}\lambda_i - \lambda_j^{1+\varepsilon})\vert a_{ij}\vert^2=\sum_{i,j} (\lambda_i - \lambda_j)\vert a_{ij}\vert^2 = 0,
\end{equation}
however, numerical analysis is required to determine the range of eigenvalues for which this vanishing trace term is small enough to overcome the $\varepsilon$ in the denominator of Equation \ref{eq:epsilon_case}. 
Because the values of $a_{ij}$ are operator-dependent, let us illustrate the principle by simply setting $\vert a_{ij}\vert = 1$ for all $i,j$ and computing the following:
\begin{equation}\label{eq:to_plot_num_0}
    \sum\limits_{i,j} \lambda_j^{\varepsilon}(\lambda_i - \lambda_j),
\end{equation}
which we refer to as the trace term. We also compute the regularised version, incorporating the $\varepsilon$ in the denominator:
\begin{equation}\label{eq:to_plot_num}
    \frac{(1+\varepsilon)\sum\limits_{i,j} \lambda_j^{\varepsilon}(\lambda_i - \lambda_j)}{\varepsilon}.
\end{equation}
We illustrate the results of these computations for a qutrit system in Figure \ref{fig:n_1_numerical}.\footnote{We choose a qutrit rather than a qubit because in the latter case a single vanishing eigenvalue represents a pure state. Conversely, a qutrit system may have one vanishing eigenvalue without being pure. Hence, the qutrit system is the more general case.} We set $\lambda_0 = 0.5$, such that $0\leq\lambda_1\leq 0.5$ and $\lambda_2 = 1 - \lambda_0 - \lambda_1$. We first plot the trace term, Equation \ref{eq:to_plot_num_0}, on its own to illustrate that this tends to zero as $\varepsilon\rightarrow 0$, and then plot Equation \ref{eq:to_plot_num} to illustrate the range of eigenvalues for which the $\varepsilon$ in the denominator out-competes the vanishing trace term.

On the left of Figure \ref{fig:n_1_numerical}, we indeed see that when $\varepsilon$ is small, Equation \ref{eq:to_plot_num_0} $\approx 0$ for most values of $\lambda$. However, in small $\lambda$ limit this trace term demonstrates an abrupt change in value. This abrupt change, however, is not representative of a physical change in the second time derivative. Rather, the physical behaviour is represented on the right hand side, where crucially we include the $\varepsilon$ in the denominator. The right hand side of Fig \ref{fig:n_1_numerical} thus illustrates that for relatively large eigenvalues, the magnitude of the leading order change in the von Neumann entropy is small, while in the case of one or more vanishing eigenvalues, the leading order change diverges. A further example of this behaviour is given in Appendix \ref{app:extra_plot}, where a more general form of the interaction Hamiltonian $\hat{A}$ is used.


\section{Supplementary example: leading order change in von Neumann entropy}\label{app:extra_plot}

\begin{figure}[!htb]
\centering
\begin{minipage}{0.4\textwidth}
    \centering
    \includegraphics[scale = 0.55, trim={4cm 0cm 0cm 0cm}]{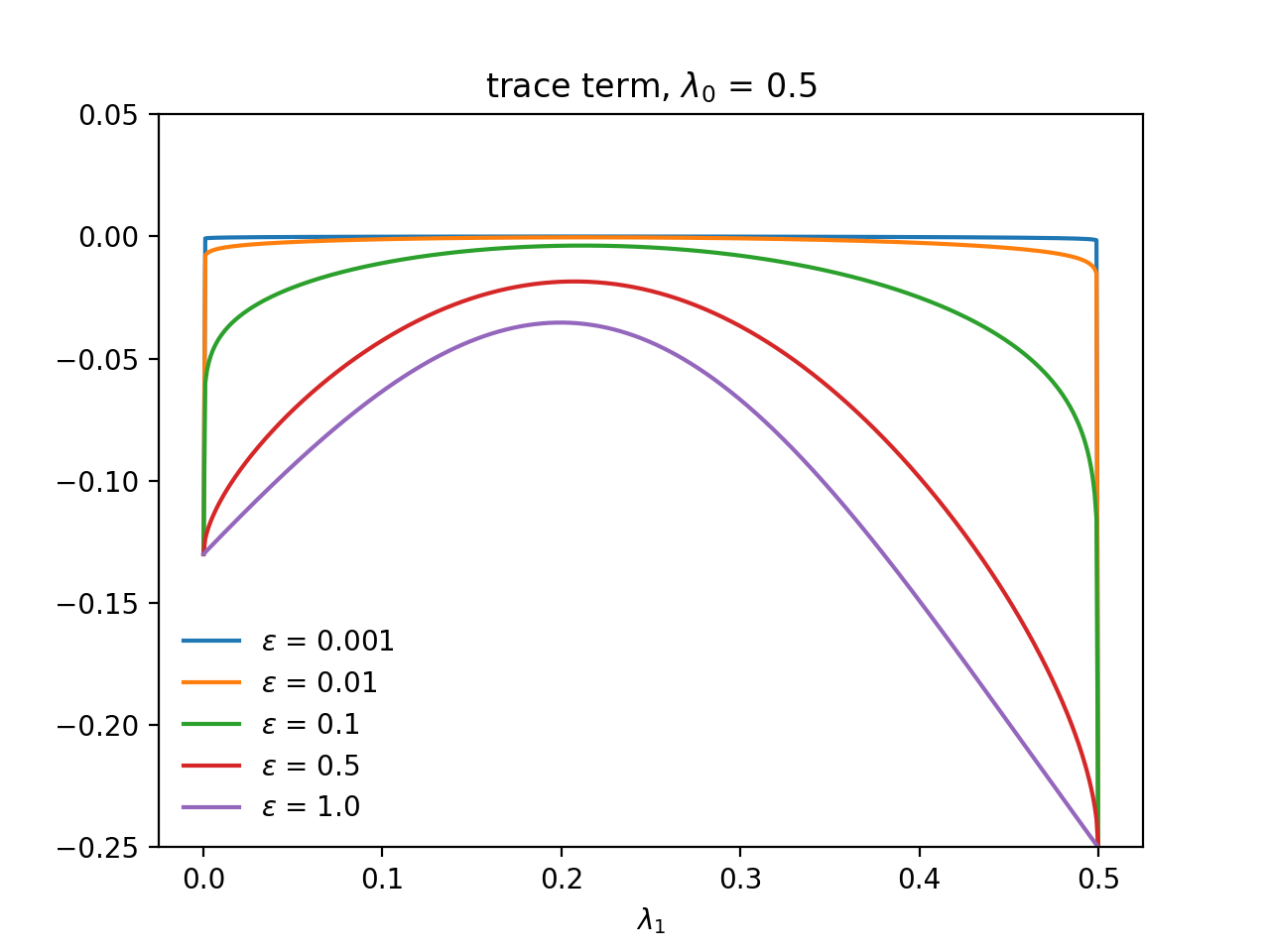}
\end{minipage}
\begin{minipage}{0.4\textwidth}
    \centering
    \includegraphics[scale = 0.55, trim={0cm 0cm 0cm 0cm}]{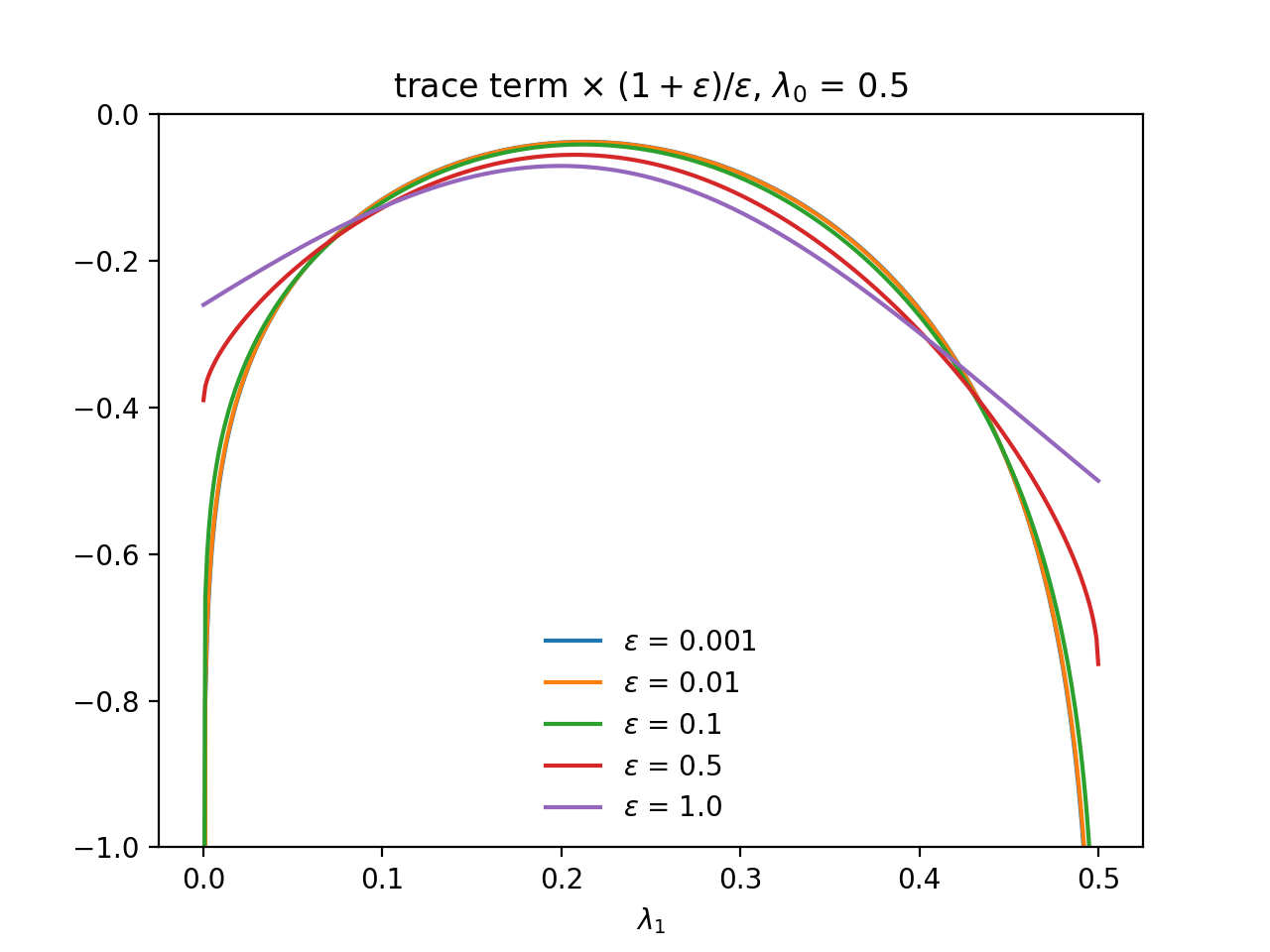}
\end{minipage}
\caption{Left: Plot of Equation  \ref{eq:trace_unreg} for a qutrit system with eigenvalue $\lambda_0 = 0.5$ and varying $\lambda_1, \lambda_2$. Right: Plot of Equation \ref{eq:reg_trace} for the same system. The interaction Hamiltonian is $\hat{A}_{\text{test}}$, resulting in an asymmetric distribution. }\label{fig:n_1_numerical_2}.  
\end{figure}

While our second order calculations may not provide clarity in terms of the ($n\rightarrow 1$) coherent information, we have demonstrated numerically that they are useful in quantifying the leading order rate of change in the von Neumann entropy of a single subsystem at the onset of an interaction. We have illustrated this numerically in Figure \ref{fig:n_1_numerical}, and provide in Figure \ref{fig:n_1_numerical_2} a further numerical example. In this latter case, we choose an interaction Hamiltonian for which $\vert a_{ij}\vert \neq 1$ for all $i,j$. Instead, we choose an arbitrary Hermitian operator as the interaction Hamiltonian, $\hat{A}_{\text{test}}$, which acts on a qutrit system with eigenvalues $\lambda_0 = 0.5$, $0\leq\lambda_1\leq 0.5$, and $\lambda_2 = 1 - \lambda_0 - \lambda_1$. The interaction Hamiltonian $\hat{A}_{\text{test}}$ is:
\begin{equation}
    \hat{A}_{\text{test}} = \begin{pmatrix}
        0.2 & 0.1 & 0.5 \\
        0.1 & 0.3 & 0.5 \\
        0.5 & 0.5 & 0.5
    \end{pmatrix},
\end{equation}
and the corresponding plots of 
\begin{equation}\label{eq:trace_unreg}
    \Tr_A\Big[\rho_A^\varepsilon\big[\hat{A}_{\text{test}}, \rho_A\big] \hat{A}_{\text{test}}\Big], 
\end{equation}
and 
\begin{equation}\label{eq:reg_trace}
    \frac{1+\varepsilon}{\varepsilon}\Tr_A\Big[\rho_A^\varepsilon\big[\hat{A}_{\text{test}}, \rho_A\big] \hat{A}_{\text{test}}\Big], 
\end{equation}
are illustrated on the left and right sides of Figure \ref{fig:n_1_numerical_2}, respectively. These quantities are related to $\ddot{H}_{1+\varepsilon}(A)\vert_{t=0}$ according to Equation \ref{eq:epsilon_case} in Section \ref{sec:n_1}. Importantly, we see here that non-trivial asymmetry can exist in the distribution of Equation \ref{eq:reg_trace} across the space of possible states of System $A$, illustrating the way in which the leading order change in the von Neumann entropy of this system at the onset of interaction is sensitive to the precise configuration of the initial input state.

\section{Further discussion of the \texorpdfstring{$n\rightarrow 1$ ($\varepsilon \rightarrow 0$)} \text{ case}}\label{app:caution}

In Section \ref{sec:n_1}, we attempted to quantify the leading (second) order change in the coherent information at the onset of a quantum channel, as illustrated in Figure \ref{fig:setup}. We found, however, that the second order change in the von Neumann entropy diverges in the case of pure states. As a consequence, the leading order change in the coherent information, given by $\ddot{S}(A)-\ddot{S}(B)$, will itself always diverge, since system $B$ is pure by definition. 

One may naturally ask whether it is possible to quantify how the speed of the divergence depends upon the state of System $A$. However, we demonstrate here that a number of issues arise when we attempt to answer this question. 

Let us revisit the expression for the leading order change in the $n$-coherent information as $n\rightarrow 1$ (or $\varepsilon \rightarrow 0$ where $n=1+\varepsilon$). This of course corresponds to the conventional definition of the coherent information, which utilises the von Neumann entropy. Representing the $n^{th}$ R\'enyi entropy as $H_n$, we have:

\begin{align}
    \delta I_{1+\varepsilon}^d &= \frac{t^2}{2}\left(\ddot{H}_{1+\varepsilon}(A)\big|_{t=0} -\ddot{H}_{1+\varepsilon}(B)\big|_{t=0} \right), \\[1em]
    &=\frac{(1+\varepsilon) t^2}{-\varepsilon}\Big((\Delta B)^2 D_{1+\varepsilon,A}-(\Delta A)^2D_{1+\varepsilon,B}\Big).\label{eq:caution}
\end{align}

Because System $B$ is pure by design, $D_{1+\varepsilon, B} = (\Delta B)^2$. Hence, we may factor out the $(\Delta B)^2$ to obtain:
\begin{equation}
        \delta I_{1+\varepsilon}^d=\frac{(1+\varepsilon) t^2(\Delta B)^2}{\varepsilon}\Big((\Delta A)^2 - D_{1+\varepsilon,A}\Big),
\end{equation}

From here, we may attempt to characterise the speed of the divergence by neglecting the prefactor of $t^2(\Delta B)^2(1+\varepsilon)/\varepsilon$, which is independent of System $A$, and simply consider the magnitude of the term $(\Delta A)^2 - D_{1+\varepsilon,A}$. However, we know from Section \ref{sec:n_1} that for $\varepsilon\rightarrow 0$, $D_{1+\varepsilon, A}$ is approximately zero across the whole state space of System $A$, except where one or more eigenvalues vanish. Consequently, we expect $(\Delta A)^2 - D_{1+\varepsilon,A}\approx (\Delta A)^2$ except at the positions in state space corresponding to vanishing eigenvalues. Much like the left side of Figure \ref{fig:n_1_numerical}, the distribution of this factor would therefore exhibit a sharp jump, approaching a discontinuity as $\varepsilon \rightarrow 0$. Such behaviour cannot be representative of a physical quantity, as this would suggest that states which were arbitrarily close to purity would exhibit vastly different behaviour to states which were exactly pure. Indeed, we showed in Section \ref{sec:n_1} that the $1/\varepsilon$ prefactor cannot be neglected if we want to obtain a smooth distribution across state space. However, if we are to incorporate this prefactor into the difference term $(\Delta A)^2 - D_{1+\varepsilon, A}$, we simply obtain a divergence as $\varepsilon\rightarrow 0$ because of the finite value of the $(\Delta A)^2$ term. Hence, we conclude that this approach is unsuitable for characterising the speed of the divergence in $\delta I^d_{1+\varepsilon}$.

One might also consider the possibility that System $B$ is not completely pure. In such a case, we could not factor out $(\Delta B)^2$, and we must instead consider the difference $(\Delta B)^2 D_{1+\varepsilon,A}-(\Delta A)^2D_{1+\varepsilon,B}$. If we were to assume that system $B$ was not completely pure, both $D_{1+\varepsilon,A}$ and $D_{1+\varepsilon,B}$ approach zero, and there appears to be a competition between the two terms which could be regularised by the factor of $1/\varepsilon$, similar to the right hand of Figure \ref{fig:n_1_numerical}. However, it is important to note that in the derivation of Equation \ref{eq:caution} we required that the tripartite system $A\tilde{A}B$ be pure. Furthermore, we required that systems $A$ and $B$ are initially unentangled such that $\rho_{AB} = \rho_A\otimes\rho_B$. Hence, we cannot simply assume that system $B$ is impure while ensuring the validity of Equation \ref{eq:caution}.

One approach, however, may be to decompose system $B$ into two subsystems $B_1$ and $B_2$, such that $B$ as a whole is pure, but the individual subsystems are not. One may then restrict the interaction Hamiltonian such that it acts trivially on one subsystem, i.e. $\hat{B} = \hat{B}_1\otimes\mathbb{I}_2$. However, this constitutes a different and more complicated physical scenario to that which we have been considering in this work. Indeed, if $B_1$ and $B_2$ are entangled, we have:
\begin{equation}
    \rho_B = \sum_{i,j,k,l} a_{ijkl}\ket{i}\bra{j}_1\otimes\ket{k}\bra{l}_2 := \sum_{i,j,k,l} a_{ijkl}\sigma_1{}_{ij}\otimes \sigma_2{}_{kl}.
\end{equation}
Because this is not a product state, we cannot simply exchange $B$ for $B_1$ in the durability expression. Instead we have:

\begin{align}
    &\Tr_B\Big[\rho_B^{n-1}\big[\hat{B},\rho_B\big]\hat{B}\Big] =\\
    &\Tr_{1,2}\Bigg[\Big(\sum_{pqrs} a_{pqrs}\sigma_1{}_{pq}\otimes\sigma_2{}_{rs}\Big)^{n-1}\Big(\sum_{ijkl} a_{ijkl}\Big[\hat{B}_1,\sigma_1{}_{ij}\Big]\hat{B}_1\otimes\sigma_2{}_{kl}\Big)\Bigg].
\end{align}

This represents an interesting problem, as it may be more plausible to consider that System $B$ is not initially completely pure. However, quantifying this slight impurity for the purpose of computing the difference term is non-trivial, and we leave further exploration of this to future work.

We anticipate that incorporating higher perturbative orders would be useful in regularising the divergences of the $n\rightarrow 1$ case. In particlar, it is likely that vanishing higher derivatives may suppress the magnitude of the entanglement transferred to the environment, even as the second derivative appears to diverge. However, we note that while the free Hamiltonians can be neglected to second perturbative order, they would need to be incorporated at higher orders. This would then significantly increase the complexity of the computations. Hence, we relegate such an approach to future work.



\end{document}